\documentstyle[epsfig,twoside]{article}
    \catcode`@=11
\begin{document}

    \thispagestyle{empty} 
    \vspace*{-1.5 true cm}


\def\Td#1{
 
\noindent
\begin{minipage}{140mm}
\begin{minipage}{95mm}
{\small  Chin. J. Astron. Astrophys. {#1}\\  
\hfill~~ \\
\hfill~~ \\
{\bf INVITED REVIEWS}
}
\end{minipage} \ \hfill  \
\begin{minipage}{35mm}
\normalsize
\begin{tabular}{l}
           \hline\\[-3mm]
          Chinese Journal of\\
          Astronomy and \\
          Astrophysics\\[1.8mm] 
\hline \\[-3mm]
\end{tabular}
\end{minipage}
\end{minipage}
} 

\def\sec@upcase#1{\relax{#1}}
\newif\if@firstsection \@firstsectiontrue
\def\section{\if@firstsection
\@firstsectionfalse\fi
\@startsection {section}{1}{\z@}
{4ex plus .5ex}{2ex plus .2ex}{\normalsize\bf}}
\def\subsection{\@startsection{subsection}{2}{\z@}
{2ex plus .5ex}{1ex plus .2ex}{\normalsize\bf}}
\def\subsubsection{\@startsection{subsubsection}{3}{\z@}
{2ex plus .5ex}{1ex plus .2ex}{\normalsize\it}}
\def\thesection{\@arabic{\c@section}}
\def\thesubsection{\thesection.\@arabic{\c@subsection}}
\def\thesubsubsection{\thesubsection.\@arabic{\c@subsubsection}}

\newcommand\acknowledgments{\vskip 3ex\@plus.8ex\@minus.4ex}%
\let\acknowledgements=\acknowledgments

\newcounter{plate}
\def\theplate{\@arabic\c@plate}
\def\fps@plate{bp}
\def\ftype@plate{4}                     
\def\ext@plate{lof}                     
\def\fnum@plate{{\bf Plate \theplate.}}
\def\plate{\@float{plate}}
\let\endplate\end@float
\@namedef{plate*}{\@dblfloat{plate}}
\@namedef{endplate*}{\end@dblfloat}
\let\platewidth=\tablewidth
\def\platenum#1{\def\theplate{#1}\let\@currentlabel\theplate
\addtocounter{plate}{\m@ne}}
\def\thefigure{\@arabic\c@figure}
\def\fnum@figure{{\rm Fig.\space\thefigure.---}}
\def\thetable{\@arabic\c@table}
\def\fnum@table{{\rm Table \thetable:}}
\def\fps@figure{bp}
\def\fps@table{bp}
\@ifundefined{epsfbox}{\@input{epsf.sty}}{\relax}
\def\eps@scaling{.95}
\def\epsscale#1{\gdef\eps@scaling{#1}}
\def\plotone#1{\centering \leavevmode
\epsfxsize=\eps@scaling\columnwidth \epsfbox{#1}}
\def\plottwo#1#2{\centering \leavevmode
\epsfxsize=.45\columnwidth \epsfbox{#1} \hfil
\epsfxsize=.45\columnwidth \epsfbox{#2}}
\def\plotfiddle#1#2#3#4#5#6#7{\centering \leavevmode
\vbox to#2{\rule{0pt}{#2}}
\includegraphics{#1}}


\def\markcite{\@ifnextchar\bgroup{\@markcite}
          {\@latexerr{Missing key on markcite command}
          {Each markcite command should have a key corresponding to a reference somewhere in the references section}}}
\def\@markcite#1{\relax}
\def\@citex[#1]#2{\if@filesw\immediate\write\@auxout{\string\citation{#2}}\fi
\def\@citea{}\@cite{\@for\@citeb:=#2\do
{\@citea\def\@citea{,\penalty\@m\ }\@ifundefined
{b@\@citeb}{\@warning
{Citation `\@citeb' on page \thepage \space undefined}}%
{\csname b@\@citeb\endcsname}}}{#1}}


\let\jnl@style=\rm
\def\ref@jnl#1{{\jnl@style#1}}
\def\aj{\ref@jnl{AJ}}                             
\def\araa{\ref@jnl{ARA\&A}}             
\def\apj{\ref@jnl{ApJ}}                           
\def\apjl{\ref@jnl{ApJ}}                
\def\apjs{\ref@jnl{ApJS}}               
\def\ao{\ref@jnl{Appl.~Opt.}}           
\def\apss{\ref@jnl{Ap\&SS}}             
\def\aap{\ref@jnl{A\&A}}                
\def\aapr{\ref@jnl{A\&A~Rev.}}                    
\def\aaps{\ref@jnl{A\&AS}}              
\def\azh{\ref@jnl{AZh}}                           
\def\baas{\ref@jnl{BAAS}}               
\def\jrasc{\ref@jnl{JRASC}}             
\def\memras{\ref@jnl{MmRAS}}            
\def\mnras{\ref@jnl{MNRAS}}             
\def\pra{\ref@jnl{Phys.~Rev.~A}}        
\def\prb{\ref@jnl{Phys.~Rev.~B}}        
\def\prc{\ref@jnl{Phys.~Rev.~C}}        
\def\prd{\ref@jnl{Phys.~Rev.~D}}        
\def\pre{\ref@jnl{Phys.~Rev.~E}}        
\def\prl{\ref@jnl{Phys.~Rev.~Lett.}}    
\def\pasp{\ref@jnl{PASP}}               
\def\pasj{\ref@jnl{PASJ}}               
\def\qjras{\ref@jnl{QJRAS}}             
\def\skytel{\ref@jnl{S\&T}}             
\def\solphys{\ref@jnl{Sol.~Phys.}}      
\def\sovast{\ref@jnl{Soviet~Ast.}}      
\def\ssr{\ref@jnl{Space~Sci.~Rev.}}     
\def\zap{\ref@jnl{ZAp}}                           
\def\nat{\ref@jnl{Nature}}              
\def\iaucirc{\ref@jnl{IAU~Circ.}}
\def\aplett{\ref@jnl{Astrophys.~Lett.}}
\def\apspr{\ref@jnl{Astrophys.~Space~Phys.~Res.}}
\def\bain{\ref@jnl{Bull.~Astron.~Inst.~Netherlands}}
\def\fcp{\ref@jnl{Fund.~Cosmic~Phys.}}
\def\gca{\ref@jnl{Geochim.~Cosmochim.~Acta}}
\def\grl{\ref@jnl{Geophys.~Res.~Lett.}}
\def\jcp{\ref@jnl{J.~Chem.~Phys.}}      
\def\jgr{\ref@jnl{J.~Geophys.~Res.}}    
\def\jqsrt{\ref@jnl{J.~Quant.~Spec.~Radiat.~Transf.}}
\def\memsai{\ref@jnl{Mem.~Soc.~Astron.~Italiana}}
\def\nphysa{\ref@jnl{Nucl.~Phys.~A}}
\def\physrep{\ref@jnl{Phys.~Rep.}}
\def\physscr{\ref@jnl{Phys.~Scr}}
\def\planss{\ref@jnl{Planet.~Space~Sci.}}
\def\procspie{\ref@jnl{Proc.~SPIE}}
\let\astap=\aap
\let\apjlett=\apjl
\let\apjsupp=\apjs
\let\applopt=\ao
\def\phn{\phantom{0}}
\def\phd{\phantom{.}}
\def\phs{\phantom{$-$}}
\def\phm#1{\phantom{#1}}
\def\earth{\hbox{$\oplus$}}
\def\lesssim{\mathrel{\hbox{\rlap{\hbox{\lower4pt\hbox{$\sim$}}}\hbox{$<$}}}}
\def\gtrsim{\mathrel{\hbox{\rlap{\hbox{\lower4pt\hbox{$\sim$}}}\hbox{$>$}}}}
\def\sq{\hbox{\rlap{$\sqcap$}$\sqcup$}}
\def\arcdeg{\hbox{$^\circ$}}
\def\arcmin{\hbox{$^\prime$}}
\def\arcsec{\hbox{$^{\prime\prime}$}}
\def\fd{\hbox{$.\!\!^{\rm d}$}}
\def\fh{\hbox{$.\!\!^{\rm h}$}}
\def\fm{\hbox{$.\!\!^{\rm m}$}}
\def\fs{\hbox{$.\!\!^{\rm s}$}}
\def\fdg{\hbox{$.\!\!^\circ$}}
\def\farcm{\hbox{$.\mkern-4mu^\prime$}}
\def\farcs{\hbox{$.\!\!^{\prime\prime}$}}
\def\fp{\hbox{$.\!\!^{\scriptscriptstyle\rm p}$}}
\def\micron{\hbox{$\mu$m}}
\let\la=\lesssim                        
\let\ga=\gtrsim
\def\case#1#2{\hbox{$\frac{#1}{#2}$}}
\def\slantfrac#1#2{\hbox{$\,^#1\!/_#2$}}
\def\onehalf{\slantfrac{1}{2}}
\def\onethird{\slantfrac{1}{3}}
\def\twothirds{\slantfrac{2}{3}}
\def\onequarter{\slantfrac{1}{4}}
\def\threequarters{\slantfrac{3}{4}}
\def\ubvr{\hbox{$U\!BV\!R$}}            
\def\ub{\hbox{$U\!-\!B$}}               
\def\bv{\hbox{$B\!-\!V$}}               
\def\vr{\hbox{$V\!-\!R$}}               
\def\ur{\hbox{$U\!-\!R$}}               
\def\ion#1#2{#1$\;${\small\rm\@Roman{#2}}\relax}
\def\nodata{ ~$\cdots$~ }
\newcount\lecurrentfam
\def\LaTeX{\lecurrentfam=\the\fam \leavevmode L\raise.42ex
\hbox{$\fam\lecurrentfam\scriptstyle\kern-.3em A$}\kern-.15em\TeX}
\def\sizrpt{
(\fontname\the\font): em=\the\fontdimen6\font, ex=\the\fontdimen5\font
\typeout{
(\fontname\the\font): em=\the\fontdimen6\font, ex=\the\fontdimen5\font
}}


\newcommand{\no}{\noindent}
\newcommand{\dg}{\mbox{$^\circ$}}
\newcommand{\vs}{\vspace{2.5mm}}
\newcommand{\vn}{\vs\no}
\newcommand{\hs}{\hspace{2.5mm}}
\newcommand{\hf}{\hfill}
\newcommand{\dn}[1]{\mbox{$_{#1}$}} 
\newcommand{\para}[1]{\vs\vs\bc{\bf#1}\ec}
\newcommand{\be}{\begin{equation}}
\newcommand{\ee}{\end{equation}}
\newcommand{\bc}{\begin{center}}
\newcommand{\ec}{\end{center}}
\newcommand{\sun}{\mbox{$_\odot$}}
\newcommand{\ti}[1]{\mbox{$\times 10^{#1}$}}
\newcommand{\x} {\mbox{$\times$}}
\newcommand{\bm}[1]{\mbox{\boldmath $#1$}} 
\newcommand{\fns}{\footnotesize\ }
\newcommand{\nor}{\normalsize\ }
\newcommand{\mult}{\multicolumn}
\newcommand{\rmd}{{\rm d}}

\newcommand{\rarr}{\rightarrow}
\newcommand{\larr}{\leftarrow}
\newcommand{\pll}{\parallel}  
\newcommand{\rme}{{\rm e}}
\newcommand{\mod}{{\rm mod}}
\newcommand{\au}{{\sc au}}

\newcommand{\av}[1]{\mbox{$\langle{#1}\rangle$}}
\newcommand{\alfv}{Alfv\'{e}n}
\def\dfrac#1#2{{\displaystyle{#1\over#2}}}
\def\df {\dfrac}

\newcommand{\alf}{\mbox{$\alpha$}}
\newcommand{\bet}{\mbox{$\beta$}}
\newcommand{\gam}{\mbox{$\gamma$}}
\newcommand{\dlt}{\mbox{$\delta$}}
\newcommand{\eps}{\mbox{$\epsilon$}}
\newcommand{\veps}{\mbox{$\varepsilon$}}
\newcommand{\vphi}{\mbox{$\varphi$}}
\newcommand{\lam}{\mbox{$\lambda$}}
\newcommand{\muu}{\mbox{$\mu$}}
\newcommand{\mum}{\mbox{$\mu$m}}
\newcommand{\mus}{\mbox{$\mu$s}}
\newcommand{\nuu}{\mbox{$\nu$}}
\newcommand{\sgm}{\mbox{$\sigma$}}
\newcommand{\p}{\partial}

\newcommand{\ha}{H\dn{\alf}}
\newcommand{\hb}{H\dn{\bet}}
\newcommand{\hg}{H\dn{\gam}}
\newcommand{\hd}{H\dn{\dlt}}

\newcommand{\up}[1]{\mbox{$^{#1}$}}
\newcommand{\uph}{$^{\rm h}$}
\newcommand{\upm}{$^{\rm m}$}
\newcommand{\ups}{$^{\rm s}$}
\newcommand{\hms}[3]{\mbox{#1\uph #2\upm #3\ups}}
\newcommand{\dms}[3]{\mbox{#1\dg #2$'$ #3$''$}}

\newcommand{\ksm}{\mbox{km\,s\up{-1}\,Mpc}}
\newcommand{\gcm}{\mbox{g\,cm\up{-3}}}
\newcommand{\jcs}{\,J\,cm\up{-2}\,s\up{-1}}

\newcommand{\cas}{The Chinese Academy of Sciences}

\newcommand{\bao}{{\it Beijing Astronomical Observatory, 
The Chinese Academy of 
          Sciences, Beijing 100012}}

\newcommand{\pmo}{{\it Purple Mountain Observatory, The Chinese Academy of 
          Sciences, Nanjing 210008}\\}

\newcommand{\sho}{{\it Shanghai Observatory, The Chinese Academy of 
          Sciences, Shanghai 200030}}

\newcommand{\yuo}{{\it Yunnan Observatory, The Chinese Academy of 
          Sciences, Kunming 650011}}
\newcommand{\ust}{{\it Centre for Astrophysics, University of 
          Sciences \& Technology of China, Hefei 230026}\\}
\newcommand{\ess}{{\it Department of Earth and Space Sciences, University of 
        Sciences \& Technology\\ of China, Hefei 230026}\\}
\newcommand{\gra}{{\it Graduate School, University of Sciences 
          \& Technology of China, Beijing 100039}\\}
\newcommand{\nua}{{\it Department of Astronomy, Nanjing University, 
          Nanjing 210008}\\}
\newcommand{\nnup}{{\it Department of Physics, Nanjing Normal University,
          Nanjing 210097}\\}

\newcommand{\nup}{{\it Department of Physics, Nanjing University, 
          Nanjing 210008}\\}
\newcommand{\geo}{{\it Department of Geophysics, Peking University, 
          Beijing 100871}\\}
\newcommand{\phy}{{\small\it Department of Physics, Peking University 
         }}
\newcommand{\hep}{{\it Institute of High Energy Physics, Chinese 
          Academy of Sciences, Beijing 100039}\\}
\newcommand{\itp}{{\it Institute of Theoretical Physics, Chinese 
          Academy of Sciences, Beijing 100080}\\}
\newcommand{\bnua}{{\it Department of Astronomy, Beijing Normal 
          University, Beijing 100875}\\}
\newcommand{\bnup}{{\it Department of Physics, Beijing Normal 
          University, Beijing 100875}\\}
\newcommand{\sxo}{{\it Shaanxi Observatory, The Chinese Academy of 
          Sciences, Lintong 710600}\\}
\newcommand{\jlos}{{\it Joint Laboratory for Optical Astronomy, 
         The Chinese Academy of Sciences, Shanghai 200030}\\}
\newcommand{\jlrn}{{\it Joint Laboratory for Radio Astronomy, 
        The Chinese Academy of Sciences, Nanjing 210008}\\}
\newcommand{\css}{{\it Center for Space Science and Applied
        Research, Beijing 100080}\\}
\newcommand{\rajl}{{\it Radio Astronomy Joint Laboratory, Beijing, 
        100080}\\}
\newcommand{\urum}{{\it Urumqi Astronomical Station, Urumqi,
        830011}\\}
\newcommand{\jtap}{{\it Department of Applied Physics, Shanghai Jiaotong
        University, Shanghai 200240}\\}
\newcommand{\ispa}{{\it Institute of Space Physics and Astrophysics, 
        Shanghai Jiaotong Univeristy, Shanghai 200030}\\}


\newcommand{\jbac}{{\it CAS-Peking University Joint Beijing 
        Astrophysical Center, Beijing 100871}\\}

\newcommand{\nao}{{\it National Astronomical Observatories, 
        The Chinese Academy of Sciences, Beijing 100012}}
\newcommand{\hua}{{\it Department of Physics, Huazhong University of 
         Science and Technology, Wuhan 430074}\\}
\newcommand{\iam}{{\it Institute of Applied Mathematics, Chinese Academy of
         Sciences, Beijing 100080}\\}
\newcommand{\heb}{{\it Department of Physics, Hebei Normal University, 
         Shijiazhuang 050016}\\}



\Td{Vol. 2 (2002), No. 4,   293--324} 

\setcounter{page}{293} 

\renewcommand{\baselinestretch}{1.14}
\newcommand{\etal}{ et al. }

\vspace{10mm}
\pagestyle{myheadings} 
\markboth{{\protect\small
J. L.  Han \& R. Wielebinski
\hfil}}
{{\hfil\protect\small
Milestones in the Observations of  Cosmic Magnetic Fields
}}

\arraycolsep 1mm

\baselineskip 6mm

\vs\vs
\no
{\Large\bf
Milestones in the Observations of Cosmic Magnetic Fields
}
\footnote{
$\star$ E-mail: hjl@bao.ac.cn
}

\baselineskip 5mm

\vs
\no
{\large
Jin-Lin Han\,$^{1~ \star}$ and Richard Wielebinski\,$^2$ 
}

\baselineskip 4.5mm

\vs
\no
$^1$   
National Astronomical Observatories, Chinese Academy of Sciences, 
             Beijing 100012 

\no
$^2$
Max-Planck-Institut  f\"ur Radioastronomie, Auf dem H\"ugel 69, 
               D-53121, Bonn, Germany

\vs
{\small Received 2002 June 10 ; accepted 2002 June 14}

\begin{quotation}
\vs\vs\no {\bf Abstract}~~
Magnetic fields 
are observed everywhere in the universe.  
In this review, we concentrate on 
the observational aspects of
the magnetic fields of Galactic and extragalactic objects. Readers
can follow the milestones 
in the observations of
 cosmic magnetic fields obtained from
the most important tracers of magnetic fields, namely, the star-light
polarization, the Zeeman effect, the rotation measures (RMs,
hereafter)  of extragalactic radio sources, the pulsar RMs, radio
polarization observations, as well as 
the newly implemented sub-mm and mm
polarization capabilities.\\
The magnetic field of the Galaxy was first discovered in 1949
by optical polarization observations. The local magnetic
fields 
within one or two kpc have been well delineated by
starlight polarization data. The polarization observations of
diffuse Galactic radio background emission in 1962 confirmed
unequivocally the existence of a Galactic magnetic field.
The bulk of the present information about the magnetic fields
in the Galaxy comes from analysis of rotation measures of
extragalactic radio sources and pulsars, which can be used to
construct the 3-D magnetic field structure in the Galactic
halo and Galactic disk.  Radio synchrotron spurs in the
Galactic center show a poloidal field, and the polarization
mapping of dust emission and Zeeman observation in the
central molecular zone reveal a toroidal magnetic field
parallel to the Galactic plane. For nearby galaxies, both
optical polarization and multifrequency radio polarization
data clearly show the large-scale magnetic field following
the spiral arms or dust lanes. For more distant objects,
radio polarization is the only approach available to show the
magnetic fields in the jets or lobes of radio galaxies or
quasars. Clusters of galaxies also contain widely distributed
magnetic fields, which are reflected by radio halos or the RM
distribution of background objects. The intergalactic space
could have been magnetized by outflows or galactic superwinds
even in the early universe.  The Zeeman effect and
polarization of sub-mm and mm emission can be used for the
study 
of magnetic fields in some Galactic molecular clouds
but it is observed only at high intensity. Both approaches
together can clearly show the role that magnetic fields play
in star formation and cloud structure, which in principle
would be analogous to galaxy formation from protogalactic
clouds. The origin of the cosmic magnetic fields is an active
field of research. A primordial magnetic field has not been
as yet directly detected, but its existence must be
considered to give the seed field necessary for many
amplification processes that have been developed. Possibly,
the magnetic fields were generated in protogalactic plasma
clouds by 
the dynamo process, and maintained again by the dynamo
after galaxies were formed.

\vs\no
\end{quotation}


\section{AT THE BEGINNING}

Magnetic fields have been observed everywhere in the universe.  
Historically, the existence of the magnetic field of the
Earth has been known in China for over 4000 years. Chinese emperors
of the Han Dynasty used `magnetic carts' to point the way on the
tours of their empire. 
The Roman chronicler Plinius recorded that
in 
ancient Greece, in the province called
`Magnesia', 
iron ore with magnetic properties 
had been mined
possibly for thousands of years. The use
of the compass for navigation was practised by Chinese, Arab,
Portuguese, Spanish and English seafarers.  In England William
Gilbert performed experiments with magnets around 1600. A series of
important discoveries of magnetic effects 
were made by  Coulomb,
Faraday, Oersted, and Gauss.

The basic experiment of Zeeman, the observation of the splitting of
spectral lines by passage through a magnetic field, opened the way to
remote sensing of magnetic fields. A few years after
Zeeman's result was  published, in 1908, G. E. Hale observed magnetic
fields in the Sun. The detection of 
magnetic fields in Ap stars was
made by Babcock (1947).
 
The first discussion about 
the need 
of 
an interstellar magnetic field 
to
explain the isotropy of 
the cosmic radiation was given by Alfv\'en
(1937), but Fermi (1949) stressed that
the fields 
filled 
the vast 
expanses of interstellar space. The
first observations of the polarization of starlight, made by Hiltner
(1949) and Hall (1949), were at first
interpreted to be due to the scattering 
by dust in the
Galactic plane. An alternative explanation by Davis \& Greenstein
(1951) was that magnetic fields may align the dust
grains.

A spectacular result was the study of the polarization of the Crab
Nebula by Dombrovsky (1954) who followed up the suggestion by
Shklovskij (1953) that the light of the Crab nebula is due to
synchrotron emission and is 
therefore polarized.  Oort
\& Walraven (1956) and Woltjer (1957)
followed up these observations
with two dimensional vector plots
of the polarization of the Crab nebula,
proving that it was 
indeed 
optical synchrotron radiation. New optical polarization observations
and 
historical observations of this object can be found in
Hickson \& van den Bergh (1990).
 
The advent of radio astronomy, starting in the 1950's, allowed the
measurement of magnetic fields in a variety of cosmic objects. Data have
been gathered on the magnetic fields in the Milky Way, nearby
galaxies, clusters of galaxies and distant radio galaxies.  In this
review, 
we will not be dealing with 
the magnetic fields of the Earth, the Sun and the planets.
Instead, we will concentrate on the many aspects of
the 
Galactic and extra-galactic magnetic fields.

The detection of radio polarization 
in the Galaxy (Westerhout et al.
1962; Wielebinski et al. 1962) gave the final proof that magnetic
fields exist in the Milky Way. These early observations showed that
the radio polarization was subject to ionospheric Faraday rotation at
lower radio frequencies. Polarization of the radio continuum emission
of discrete radio sources was observed by Mayer et al. (1962). In the
succeeding years, both the Galactic polarized emission and the
discrete radio sources were shown to be subject to Faraday rotation
by
the Galactic magnetic fields along the line of sight. This discovery
gives the second radio method of measuring magnetic fields.

The Zeeman effect was used at optical wavelengths to measure the
magnetic field of the Sun and 
of the magnetic stars. The radio Zeeman effect
was predicted for the HI line by Bolton and Wild (1957) but it took
some time to be detected. The radio Zeeman effect was finally found
by Verschuur (1968) in HI clouds in the Galaxy. Zeeman effect studies
of other molecular species that probe magnetic field strengths in
different molecular clouds were also successful. The observations,
however, are difficult and data gathering during the past years
has been slow.

Soon after the discovery of the first pulsar it was realized that
these objects could be ideally used to measure the magnetic fields of
the Galaxy. This was pointed out first by Lyne \& Smith (1968). In
some ways pulsar studies combined with Galactic polarization
observations are the best method to investigate the Galactic magnetic
field.
 
The earliest maps of radio galaxies showed that considerable
polarization is present and that magnetic fields are involved
in the emission process (e.g., Cooper \& Price 1964; Morris et al.  1964;
H\"ogbom \& Carlsson 1974).  The advent of large synthesis radio
telescopes in Westerbork, Holland and the Very Large Array in
Soccoro, New Mexico contributed significantly to this subject.
 
The search for radio polarization in normal galaxies led to the first
detection of the regular magnetic fields in M51 by Mathewson et al.
(1972), the mapping of M31 by Beck et al. (1978) followed.  This
field of research, which requires observations at several high
frequencies, and has been dominated by the observations with Effelsberg radio 
telescope.
More recently, the combination of Effelsberg single dish maps with
the VLA data 
greatly advanced 
our knowledge of galactic magnetic fields.

Diffuse radio continuum emission was found to be present in clusters
of galaxies (e.g., Ryle \& Windram 1968; Willson 1970; Wielebinski 1978).
In particular, the investigation of the
Coma A cluster led to the conclusion that magnetic fields of B $\sim$
2\,$\mu$G are present in the intracluster medium between the galaxies. This 
result came both
from an equipartition argument for the continuum data (e.g.
Deiss et al.  1997) and Faraday rotation studies by
Kim et al. (1990). Certainly clusters of galaxies
represent the largest magnets in the Universe.

The existence of magnetic fields in the more distant Universe
has been the subject of many papers. We refer to 
the excellent review
by Kronberg (1994) who 
showed that Lyman-$\alpha$ systems
with $z = 2.0 \sim 3.0$ 
possess magnetic fields.
Additional reviews on magnetic fields in galaxies
have been made by Zeldovich et al.
(1983), Rees (1987), Wielebinski \& Krause
(1993), Beck et al. (1996),
Zweibel \& Heiles (1997) and Beck (2001).

Note that Zeeman splitting and Faraday rotation can detect the magnetic field
component along the line of sight, i.e., $B_{||}$, and are sensitive to its 
sign,
whereas 
synchrotron radiation and polarimetry
(of starlight or dust) mostly reflect on the 
field component    
perpendicular
to the line of sight, $B_{\perp}$.

\section{OPTICAL POLARIZATION}

In 
retrospect, the earliest method of tracing the magnetic fields of the 
Galaxy
was 
actually 
successful. 
The first reports (Hiltner 1949;
Hall 1949) of the polarization of starlight came simultaneously with
the suggestion that magnetic fields may align the dust
grains, 
and 
theoretical work by Davis
\& Greenstein (1951) implied that polarization was caused by 
dust
grains so 
lined-up.
However, the problem
of separating 
simple scattering effects 
from polarization
due to dust grains aligned in magnetic fields makes
the interpretation 
ambiguous. 
This led to a controversy that
continued for many years. Many optical astronomers were convinced
that only scattering was responsible for all observed polarization, 
while one can say that much of the polarization is in fact due to magnetic
alignment. In fact, other possibilities exist 
as 
to 
the 
cause 
of
polarization
of thermal dust emission, such as dust grains in 
an
anisotropic radiation 
field (e.g., Onaka 2000) or different populations 
of grains 
at different temperatures, see Goodman (1996) 
for a review on both 
the observational and theoretical 
aspects. 
 
\subsection{Polarization of Stars} 
 
Polarization of starlight can be used to detect 
magnetic fields
out to 1 or 2\,kpc from the Sun. 
In the late 1950s, a substantial 
catalog of the polarization of stars
in the northern hemisphere was 
compiled by Behr (1959). This
work was continued in the southern hemisphere by Mathewson and Ford
(1970a) who eventually presented an all-sky distribution
of starlight polarization. 
Their catalog 
includes polarization
measurements by Hiltner (1956), Hall (1958),
Behr (1959), Loden (1961), Appenzeller
(1968), and Visvanathan (1967). 
The general conclusion of this work 
is still
valid today
that the magnetic field of the Galaxy is in
general aligned along the Galactic plane.  Some additional effects
were also noted that indicated irregularities of the local field.

\vs\vs

\hfill\psfig{figure=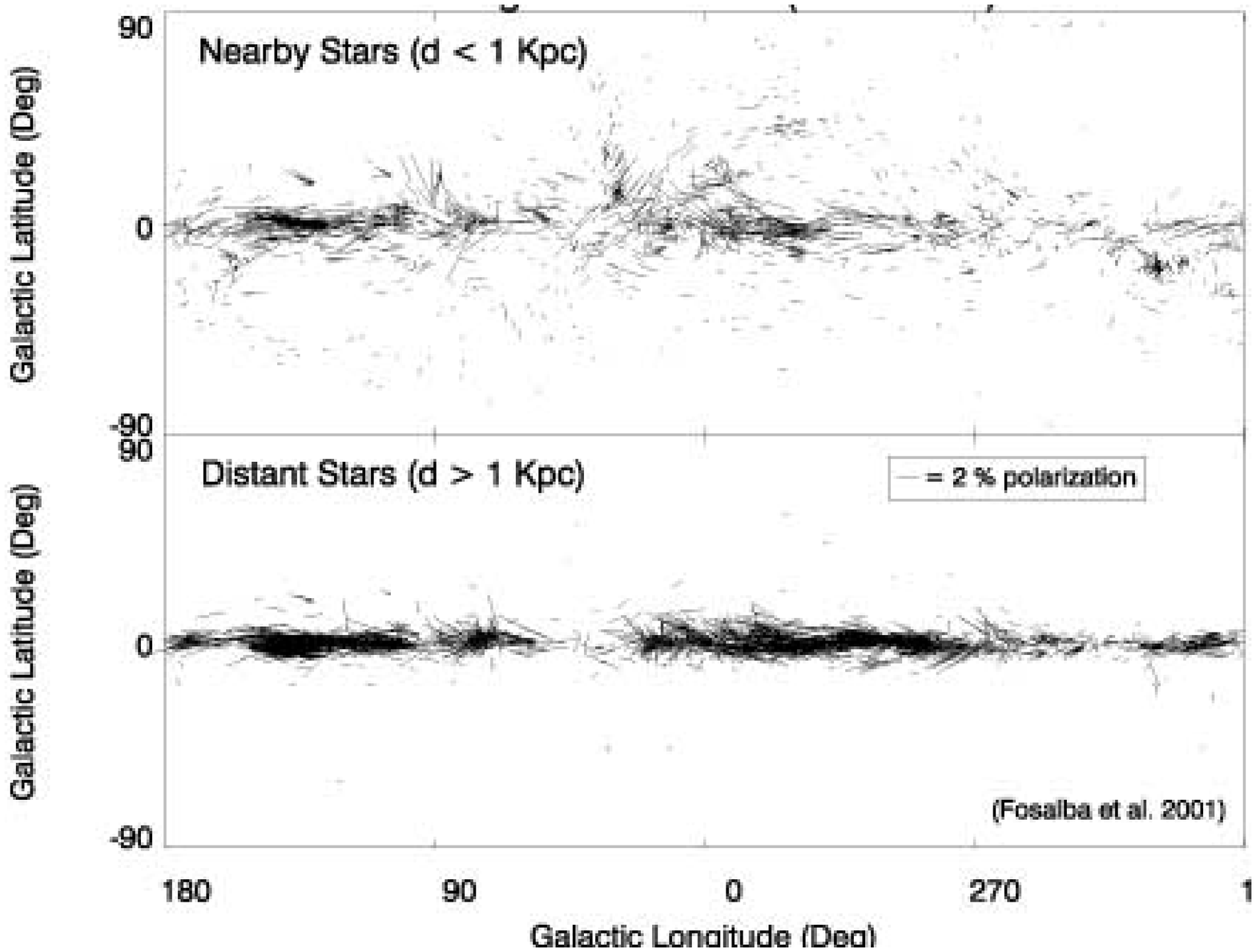, height=90mm} \hfill~~~

\bc\parbox{120mm}{\small
Fig.\,1~~
Distribution of starlight polarization. Nearby stars show
the local perturbations, and distant stars show the larger-scale field
parallel to the galactic plane (courtesy P. Fosalba).
}\ec

\vs

Using the data available 
at the time, Axon \& Ellis (1976)
compiled a catalog of 5070 stars with reliable distances.
Recently, Heiles (2000) has compiled a new catalog of
starlight polarizations of 9286 stars, using all previously
available data 
including 
1800 stars from Mathewson \&
Ford (1970a), 126 stars from Appenzeller (1974), 495 stars
from Schroeder (1976), 1660 southern OB stars from Klare \&
Neckel (1977), 313 nearby stars from Krutter (1980), 358+118
stars from Korhonen \& Reiz (1986), 1000 nearby stars from
Leroy (1993), 133  stars from Bel et al. (1993), 51 stars at
high Galactic latitudes 
from Berdyugin et al. (1995), 361 stars
from Reiz \& Franco (1998), and 126 stars from Goodman
(unpublished data). After the Heiles' catalog was compiled,
some new data from Berdyugin's group (Berdyugin \& Teerikorpi
1997, 2001; Berdyugin et al. 2001) and others (e.g., Serkowski
\& Shawl 2001) have been published. 
 
The starlight data
show that 
the 
percentage 
polarization 
increases 
with 
increasing 
extinction and/or distance (Behr 1959; 
Appenzeller 1968; Fosalba et al. 2002), 
showing the Davis and Greenstein effect from interstellar 
dust grains. Many analyses have reached the conclusion that
{\it the local regular magnetic 
fields of our Galaxy point to the direction 
$l\sim 82 \dg$, and
seem 
to follow the spiral pattern very closely} (e.g., Heiles
1996; Andreasyan \& Makarov 1989),
although the local distortions can be clearly seen 
in the nearby
stars (see Fig.~1). The polarization ``vectors'' in the southern
Galactic pole (eg. Berdyugin \& Teerikorpi 2001) seem
unperturbed by any other features and hence the direction
of local fields can be seen clearly. Most recent analysis of 
the polarization data
(e.g., Fosalba et al. 2002) shows that the regular
magnetic field is about 39\% to 62\% of the total magnetic energy.

Starlight polarization was mainly used to study the local magnetic
fields (within 1 or 2\,kpc), as 
stated above. However, it is worth 
noting
some other applications of the data. For a long time, the data were used to
probe the intervening clouds (e.g., Markkanen 1979;
Gomez de Castro et al. 1997). The excess polarization
of Vega-like stars was used to statistically study  the circumstellar
material (Bhatt \& Manoj 2000). 
Angular power spectrum
analysis of the data may be used to model the Galactic polarized 
continuum emission at other wavelengths (Fosalba et al. 2002).

\subsection{Nearby Galaxies} 
 
Optical polarization observations of nearby galaxies can reveal 
the magnetic fields in the galactic disk or dust lanes.

The observations of M31 by Hiltner (1958) were shown to
require a magnetic field aligned along the major axis. A discussion 
of the polarization produced by interstellar dust in external 
galaxies was given earlier by Elvius (1951, 
1956). Polarimetric observations of several galaxies 
by Appenzeller (1967) showed optical polarization 
vectors but were interpreted to be due to reflection as well as 
interstellar absorption with a magnetic field directed along the 
spiral arms. First polarization of stars, later photoelectric
surface photometry in Magellanic Clouds (e.g., Mathewson \& Ford 
1970b, 1970c; Schmidt 1970)
gave us some real insight of the magnetic fields in these nearby galaxies. 
 
The next important development in this field is creditted to 
S.M. Scarrott who for many years delivered numerous results 
on magnetic fields using his electronographic camera methods. 
Up to 30\% optical polarization in the halo region of M82 
(Bingham et al. 1976) and strong polarization 
along the dust lanes of M104 (Scarrott et al. 1977) 
were published. The optical polarization 
in most cases indicated the presence of large-scale magnetic 
fields in galaxies (Elvius 1978). 
The advent of 
the CCD has added more sensitivity to various
polarimeter systems. Scarrott et al. (1987) showed the
perpendicularity between the radio and optical polarization
vectors, exactly as expected if magnetic dichroism and the
synchrotron process are responsible for the polarization at
the shorter and longer wavelengths respectively. The
polarization map  of NGC\,1068 (Scarrott et al. 1991, see Fig.\,2) is so
impressive that 
apart from 
the immediate nuclear zone, the
orientation of the polarization vectors form a spiral pattern
in both the arm and interarm regions, rather than a circular
one expected from the scattering of bright nuclei. The
pattern can be explained in terms of a magnetic field with a
spiral configuration and 
by polarization produced by dichroic
extinction (see discussion by Wood 1997). Therefore, in
galaxies with low inclinations, the polarization vectors are
coherent on kpc scale-lengths and follow spiral
configurations. In galaxies with high inclinations, such as
M104 and NGC~5128 (Scarrott et al. 1996), the polarization in
the central regions is oriented parallel to the dust lane. In
NGC~1808 (Scarrott et al. 1993), the polarization
orientations follow the spiral arms on kpc scales, indicative
of a magnetic field coherent on a galactic scale (Scarrott
1996).

\vs\vs

\hfill\psfig{figure=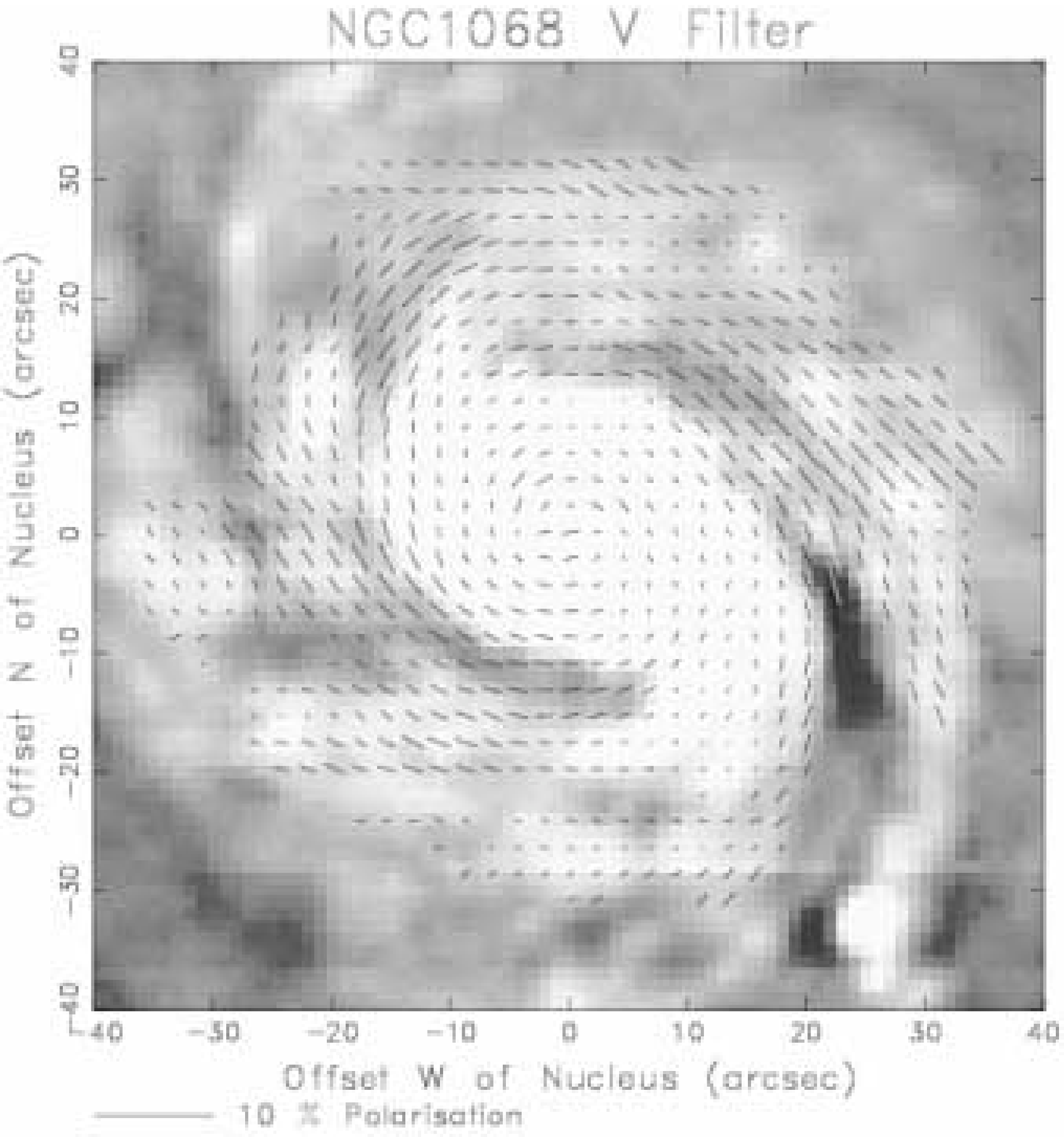, height=110mm} \hfill~~~

\bc\parbox{100mm}{\small
Fig.\,2~~ Optical polarization vectors for NGC~1068 (courtesy 
Peter W. Draper after Scarrott et al. 1991).
}\ec 

\vs

Meanwhile, other groups also contributed to the optical polarization 
of galaxies.  King (1986) found that the scattering 
of light from the bright nucleus of NGC~7331 is predominant along 
the major axis of the galaxy, elsewhere 
the polarization is consistent
with being produced by the transmission of light through elongated 
grains partially aligned by the galactic magnetic field and the 
Davis-Greenstein mechanism.  Fendt et al. (1996) 
observed three edge-on galaxies (NGC~891, 5907 and 7331). Except the 
dominant orientation given by the anisotropic scattering, NGC~891 
has polarization 
orientations indicating 
magnetic fields 
perpendicular to the major axis, and NGC~7331 has polarization 
indicating a toroidal field. These polarization maps are 
understandable after the two effects are considered and modelled 
(Wood \& Jones 1997). The polarization map of NGC~6946 
(Fendt et al. 1998) is affected  by 
foreground galactic
scattering in one 
quadrant due to its low Galactic latitude
($b\sim11.7\dg$). 
 
Optical polarization observations of more distant galaxies
(e.g., Tadhunter et al. 1992; Scarrott et al. 1990) or active
galaxies (Draper et al. 1993; Cimatti et al. 1993) have been
used to investigate some interesting issues, such as the
``unified'' scheme of AGNs, emission mechanisms (Breeveld \&
Puchnarewicz 1998) and hot spots (L\"ahteenm\"aki \& Valtaoja 
1999). One may have noticed that the Hubble Space Telescope
is equipped with polarizers, 
which provide high 
resolution polarization maps even at the 
ultraviolet band 
(e.g., Kishimoto et al. 2002),
that can be used to study 
hidden 
galactic nuclei and 
the scattering grains. 
 
\subsection{Polarization Observations at Infrared, Far-infrared, 
Millimeter, Submillimeter Ranges}

Recently  a return to the dust polarization 
method 
of
measuring 
magnetic fields has 
been made
possible 
by the advent of 
polarization measurements 
in 
the infrared, 
far-infrared, sub-mm and mm (e.g., Cudlip et al. 
1982; Dowell et al. 1998; Hildebrand et al. 2000). 
We expect no scattering at these wavelengths and 
hence the polarization originates 
in 
the emission of dust
particles 
aligned by 
the magnetic field. Observations of linear polarization from 
the thermal emission of magnetically aligned dust grains provide a 
relatively easy 
means
of 
exploring the magnetic field morphology 
(Heiles et al. 1993). Polarization observations of 
molecular clouds have been used to study the role of magnetic fields 
in the formation and evolution of the clouds and in the process of star 
formation. 
 
With greatly improved sensitivity and 
increased 
number of measuring pixels, 
it is 
now possible 
to use polarimetry to trace magnetic fields not 
only in 
the prestellar cores of bright molecular clouds  (e.g., Ward-Thompson 
et al. 2000), but also in thermal streamers, 
in the accretion disk around T Tauri stars (e.g., Tamura et al. 
1999), 
in the envelopes of young stellar objects 
(Holland et al. 1996),  and 
in other sources of 
dust emission (e.g., Vall\'ee et al. 2000). 
Magnetic fields around 
protostars (Greaves \& Holland 1998) 
are banded and there is probably a centrally contracted field, which 
collimates the bipolar outflowing gas (Greaves et al. 2001; 
Momose et al. 2001). The fields are most likely 
to be 
toroidal in the circumstellar disk (Tamura et al. 1999). 
 
On the global scale, 
polarization maps at different 
wave bands reveal the 
emission from dust grains orientated by the same magnetic fields, 
giving 
roughly consistent
results,
e.g.,  
the 
polarization maps of the Orion clouds at 1.3 mm and 3.3 mm by 
Rao et al. (1998) and 350\,$\mu$m by Hildebrand et al. 
(2000). However, one should note that the emission 
at longer wavelengths comes from 
colder dust and that at shorter 
wavelengths, 
from warmer dust. As these clouds have a temperature gradient from the 
a hot core to 
a cold ambient 
envelope or even colder large-scale clumpy 
ridges, so one should be careful 
when interpreting 
the polarization 
observations in terms of 
predominant 
magnetic fields
in 
given 
regions.
Sometimes separation is possible (e.g., Schleuning et al. 
2000). Near 
the hot 
cores or HII regions the fields 
tend to be distorted by star formation or expansions so that they 
have smaller 
length scales or 
vary rapidly in space. The inclination or 
3D structure of 
the magnetic field in some cases can be obtained by 
comparison of 
the Zeeman splitting measurements (see Schleuning et al. 
2000; Hildebrand et al. 2000). 
 
Due to limited sensitivity and resolution, it is still too
early to observe the dust polarization at these bands in
nearby Galaxies, although Greaves et al. (2000) have reported
results on M82 and Siebenmorgen et al. (2001) on NGC~1808.
The integral FIR polarization observed for the first time
from the core of the quasar 3C 279, is a completely different
story, being
related to synchrotron radiation (Klaas et al. 1999).
However, in our Galaxy, there is a ``central molecular zone''
at 
the Galactic center, 
with
a size
as large as 400 $\times$ 75 pc
(Pierce-Price et al. 2000). The thermal dust continuum
emission traces the temperature-weighted column density of
dust grains and wide-ranging network of dusty filaments, 
which can be observed at mm, sub-mm, far-infrared bands.
Werner et al. (1988) first detected the polarization in
the
far-infrared band, and concluded that the field was
predominantly azimuthal. Afterwards, more and better data
were accumulated for the circumnuclear disk (or the so-called
dust ring) in a few pc from Sgr~A* (Hildebrand et al. 1990, 
1993; Novak et al. 2000), the giant molecular cloud Sgr B2 at
100\,pc from the center as well as some other clouds near the
center (Novak et al. 1997, 2000), and for thermal filaments
(Morris et al. 1992). All 
results are consistent with a
large-scale 
azimuthal 
field with respect to the Galaxy.
Evidently the gravitational force dominates 
the magnetic force  
within the neutral gas layer and then the differential
rotation shears the field out into the azimuthal
configuration (Morris 1998).  Recent polarization
observations at 450 $\mu$m in a large area revealed the
toroidal magnetic fields 
probably 
permeating 
most of the
molecular zones ($>170$\,pc $\times$ 30\,pc) and the fields are
parallel to the Galactic plane (see Fig.6, Novak et al. 2002,
2000).  This is complementary to the dipole field delineated
by the non-thermal filaments (see next Section). 
 
\section{THE ZEEMAN EFFECT} 
 
Zeeman 
measurements are used mainly to detect the magnetic fields 
in molecular clouds and provide information about the direction of the 
field along the line of sight (see Crutcher 1999), 
which is a key step to 
understanding 
the 
process
of 
star formation 
in  
cloud 
cores (e.g., Li 1998). Combination of results from 
both polarimetry of dust emission and Zeeman observations  can 
provide a three-dimensional view of 
the 
magnetic fields in the clouds (Houde et al. 2002). 
 
As mentioned in the Introduction, the optical Zeeman effect
was used by G.E. Hale at the turn of the 
20th
century  to measure
the magnetic field of the Sun and by Babcock (1947) to
measure the magnetic fields in peculiar magnetic stars. 
A 
very important 
use of the Zeeman splitting 
is
the measurements
of spectral lines at radio frequencies. The detection of the
HI line in 1951 led to the suggestion by Bolton and Wild
(1957) to search for the Zeeman effect at radio frequencies.
This observation has turned out to be rather difficult, 
because 
the side-lobe effect and other instrumental corrections 
have 
to be
considered. Finally Verschuur (1968) and Devies et al. (1968)
gave a definite proof of the Zeeman effect in the Perseus
spiral arm by observing HI in {\it absorption} against 
Cassiopea A.  
Meanwhile, 
observation 
of 
the Zeeman effect 
in HI
{\it emission} profiles 
has been always very difficult: 
the 
inherently weak effect
due 
to the weak magnetic
fields (a few $\mu$G along the line of sight) 
is difficult 
to pick up 
against strong 
instrumental effects (e.g., Verschuur 1995a), and 
such
observations rarely
produced consistent results (e.g., Verschuur 1995b; Troland \&
Heiles 1982; Myers et al. 1995).

OH maser 
occurs in much denser clouds ($10^7$cm$^{-3}$) than diffuse 
HI emission ridges ($10^3$cm$^{-3}$), where 
the fields are also 
much stronger which helps 
the detection of Zeeman splitting. Early
attempts to identify Zeeman pairs in OH maser spectra
were 
generally not successful. Once interferometric maps
became available, one usually 
finds that 
the proposed Zeeman 
components came from widely separated regions in 
a same maser
source. 
Verification of OH maser sources came from VLBI
observations (Reid et al. 1980; Zheng et al. 2000) which
showed some pairs 
to come
from the same position within 
the 
size
of
a maser spot. 
OH masers have now been detected in 
the prestellar cores of
clouds, L1544 (Crutcher \& Troland 2000). Up till now, the
number of detections is not very 
large. All previous
detections of the OH lines (and other lines or absorptions)
with Zeeman effect have been summarized by Crutcher (1999)
and new observations 
made by, for example, Bourke et al.
(2001). 
 
The first attempt to use the Zeeman data 
to derive 
the 
large-scale magnetic field of the Milky Way was made by Reid \& 
Silverstein (1990), and later extended by Caswell \& Vaile 
(1995). It was not possible to determine 
the large-scale magnetic field of the Galaxy in view of the small 
number of sources, so a survey of 
a large number of OH masers is needed 
(see Argon et al. 2000). However, one should note that 
the star-formation activity may have changed the initial geometry of 
the magnetic field preserved during the contraction from interstellar 
density (1\,cm$^{-3}$), through 
the
density of giant molecular clouds (10$^3$\,cm$^{-3}$), to 
the 
density of OH masers near 
newly formed OB stars ($10^7$\,cm$^{-3}$). 
 
Magnetic fields of several tens of mG in interstellar H$_2$O
maser clumps ($\sim$10$^{10}$\,cm$^{-3}$) were first detected
by Fiebig and G\"usten (1989). The VLA and VLBA have been
used to study the Zeeman effect in water masers (Sarma et al.
2001). More recently attempts to detect the Zeeman effect in
a variety of lines have been made: the H30$\alpha$
recombination line by Thum and Morris (1999), the CN line by
Crutcher et al. (1999), and the CCS line by Levin et al.
(2001). Note that OH and H$_2$O masers arise in very small
regions in 
high-density clouds
under special conditions.
New detections of CN Zeeman effects offer a good chance to
measure the magnetic fields in clouds with densities of 
$10^5$ to $10^6$\,cm$^{-3}$ (Crutcher et al. 1999).

Note that observation of HI absorption towards    
the nuclear region of other (preferably edge-on) galaxies can
provide information of magnetic fields in the redshifted
clouds or circumnuclear ring around the nuclear 
region (e.g., Sarma
et al. 2002). This seems to be useful but is limited to just
a few lines of sight.

\section{THE RM of EXTRAGALACTIC SOURCES} 
 
Following the detection of the polarization of extra-galactic sources, 
the Faraday rotation effect was found by Cooper \& Price (1962) 
from the wavelength dependant emission in the source Centaurus A.
Soon afterwards, it was noticed that the rotation 
measures were dependant on both Galactic latitudes (Garder \& Whiteoak 
1963) and longitudes (Morris \& Berge 1964). 
From the emission 
source to the observer, the Faraday rotation can 
be expressed as 
\begin{equation} 
{\rm RM} = 0.810 \int_{\rm source}^{\rm Sun} n_e {\bm B} \cdot d{\bm l}, 
\end{equation} 
where 
the rotation measure, 
RM, 
is in rad~m$^{-2}$, $n_e$ is the electron density (in cm$^{-3}$), 
${\bm B}$ is the vector 
magnetic field (in $\mu$G), and $d {\bm l}$ 
is the 
line element vector of the line of sight (in pc) pointing towards us. 
The dependence implies 
that most RMs originate from (the local arm of) 
our Galaxy (see Berge \& Seielstad 1967). 
 
Studies of 
the Faraday rotation across the Galaxy were 
first made 
by Gardner \& Davies (1966) and later by Gardner et al. (1969). 
Many observers continued to gather data on the 
polarization of discrete sources (e.g., Kronberg and his students: 
Vallee \& Kronberg 1975; Kronberg \& Wardle 1977), 
however the process of collecting (reliable?) data
in early days
 is slow
(e.g., 
Tabara \& Inoue 1980). 
The successful polarization observation of radio source  led to 
RM catalogs 
across the sky (e.g., Ellis \& Axon 1978; 
Simrad-Normandin et al. 1981; Broten et al. 1988). 
A fit of RMs to a model that included a longitudinal (azimuthal) 
magnetic field with a local anomaly was made by Vallee \& Kronberg 
(1975). Many papers with a re-analysis of the existing 
data followed (e.g., Ruzmaikin \& Sokoloff, 1977; Sofue et al. 1979; 
Inoue \& Tabara 1981). Simard-Normandin \& 
Kronberg (1980) took the average RM of a cone 15$\dg$ 
radius to represent the Galactic RM sky and for the first time identified 
several large features related to loop II, loop I, etc. They showed 
that the RM at high latitudes is well correlated with,
and influenced by 
the large scale prevailing magnetic fields. This 
eventually became a 
standard for discussing ``the Galactic Rotation Measure''. 
 
  
\vs\vs\vs

\hfill\psfig{figure=ers.ps,width=60mm,angle=270} \hfill~~~

\bc\parbox{100mm}
{\small
Fig.\,3~~ 
Antisymmetric RM sky, shown by extragalactic radio sources. 
Filled symbols represent 
positive RMs, and open symbols,  
negative RMs 
(after Han et al. 1997). 
}\ec 
 
\vs
The antisymmetric RM sky (see Fig.\,3) 
was first noticed by Berge \& Seielstad 
(1967) who even 
interpreted that as a result of 
magnetic fields directed toward $l=260\dg$ for $b>0\dg$ and 
toward $l=80\dg$ for $b<0\dg$. The antisymmetric feature was 
confirmed later by Vallee \& Kronberg (1975) and Simard-Normandin 
\& Kronberg (1980) with much more data, but was 
interpreted as the local perturbations of loop I. Andreasyan (1980) 
attributed the antisymmetry to the local disk 
field with reversed 
directions above and below the plane, as Berge \& Seielstad (1967) 
suggested, while Han et al. (1997) argued that the 
antisymmetry is the result of the magnetic field in the thick 
disk or halo 
on the Galactic scale. Coincident with the vertical 
fields in the Galactic center, they suggested that A0 dynamo is 
responsible for the field structure. 
 
Reliable RMs are now available for some 1000 sources in the
literature. The sources are not well distributed on the sky.
The main weakness is that very few sources with known RM are
available  within $|b|<10\dg$ where the Galactic magnetic field is
concentrated. Only a few observations at low latitudes (Clegg
et al. 1992) produced good measurements. The situation will 
soon be improved significantly by the newly determined RMs of
point sources from the Galactic Plane polarization surveys in
the Southern and Northern 
skies (Gaensler et al. 2001; Brown \&
Taylor 2001). Nevertheless, Simard-Normandin \& Kronberg
(1980) and Sofue \& Fujimoto (1983) compared the RM data at
low latitudes with model predictions and found that the spiral
fields with direction reversals are more suitable for our
Galaxy. Concentrating on selected Galactic regions, Broten et
al. (1988) have confirmed the existence of magnetic field
reversals in the Milky Way. The most recent analysis by Frick
et al. (2001) using wavelet method basically confirmed the
previous conclusions on field reversals and the antisymmetry.
 
The RMs of extragalactic radio sources have also been used to
investigate the small-scale (200\,pc -- 0.01\,pc) structures of
interstellar media and magnetic fields. This was mostly done
by Cordes' group (Simonetti et al. 1984; Simonetti \& Cordes
1986; Lazio et al. 1990; Clegg et al. 1992). They computed
the structure function of RM spatial variations and found
that the variation of ($n_e$, $B$) can be described by a
power-law spectrum of turbulence, which is enhanced near the
Galactic plane. Both the turbulent fluctuations of electron 
density (see Armstrong et al. 1995) and random magnetic
fields are responsible for the variations (see Minter \&
Spangler 1996). For most directions the $\delta$RM is less
than 10 rad~m$^{-2}$ in a few arcminutes for background
sources, but the RM difference can be up to a few tens of
rad~m$^{-2}$ near the Galactic plane, and small anomalous
regions are possibly associated with some local Galactic
features (e.g., Clegg et al. 1992). 
 
Within 
the scope of the Canadian Galactic Plane Survey (Landecker
et al. 2000), the 
RMs
of some 380 sources 
were determined 
(Brown \& Taylor 2001). All these sources are in the $b=\pm 
4\dg$ range of the Galactic plane. This is in contrast to the
data of Simard-Normandin \& Kronberg (1980)
where 
the    
sources 
are 
mostly away from the Galactic plane. In particular,
the region of  $l = 92.0\dg,\ b = 0.5\dg$, investigated in
detail by Brown \& Taylor (2001), shows that most of the
extra-galactic sources have negative rotation measures.


\section{MAGNETIC FIELDS FROM OBSERVATIONS OF PULSARS} 
 
The discovery of pulsars gave us a new method of 
measuring 
magnetic field. Since, 
for a pulsar, 
both the RM and the dispersion
measure (DM) 
can be determined, the average 
magnetic field can 
be directly obtained, if 
the electron
density and 
the magnetic field B are not correlated (Beck 2001).
This was pointed out by Lyne \& Smith (1968) and taken up by
several investigators. A significant collection of
observations of pulsar RMs was presented by Manchester (1972,
1974), who revealed the local uniform magnetic field of about
2.2$\pm$0.4 $\mu$G, and directed towards about $l\sim90\dg$.
Thomson \& Nelson (1980) re-analyzed 48 pulsar  RMs listed by
Manchester \& Taylor (1977) and found 
field reversal near
the first inner spiral arm, i.e., the Carina-Sagittarius arm. 
 
The 
acquisition 
of RMs and DMs has become an important
``industry'' 
in pulsar astronomy. A significant step 
here 
was made by Hamilton \& Lyne (1987), who
measured 
the Faraday rotation of 163 pulsars, so 
increasing the total
number of pulsar RM 
measured 
to
185. Afterwards, Costa et al.
(1991), Rand \& Lyne (1994), Qiao et al. (1995), van Ommen et
al. (1997) and Han et al. (1999) contributed a number of new
RMs, 
providing RM data 
for some 320 objects. Note that
pulsar  observations are more concentrated in the Galactic
plane than 
the 
observations of 
extragalactic sources, so pulsars are mostly used
to probe the magnetic fields in the Galactic disk. 
 
The interpretation of the pulsar RMs as a signature 
of 
galactic magnetic fields goes back to the 
papers 
of Simard-Normandin \& Kronberg (1980) and Lyne \& Graham
Smith (1989). The latter authors confirmed the local field
strength and direction found by Manchester (1974) and the
reversal near the Carina-Sagittarius arm. They also suggested
another field reversal outside the Perseus arm from a comparison
of 
the pulsar RMs with the RMs of extragalactic radio sources. 
 
Re-analyses of pulsar RM data mostly gave more detailed
modeling 
of the structure of Galactic magnetic fields.
Rand and Kulkarni (1989) noticed the significant effect of
the North Polar Spur on pulsar RMs and suggested a
concentric-ring model for the reversed fields. This was
continued by Rand \& Lyne (1994). Since the field reversals
are expected in a bi-symmetric field structure on large
scales, the model of bi-symmetric spiral field was fitted by
Han \& Qiao (1994) and Indrani \& Deshpande (1998) after
de-projecting all RMs onto the Galactic plane. This proved to
be very successful for the local region ($<3$\,kpc). New RMs
of distant pulsars by Rand and Lyne (1994) and Han et al.
(1999, 2002) suggested the second and the third
field reversals near the
Crux-Sctum arm and the Norma arm. The most recent status in
this field, described using much more new data was given by
Han et al. (2002) who reported the detection of
 counterclockwise magnetic
field near the Norma arm (see Fig.\,4). 
 

\hfill\psfig{figure=psr.ps,width=90mm,angle=270}\hfill~~~

\bc\parbox{110mm}{\small
Fig.\,4~~ 
RMs of pulsars in the Galactic disk reveal the large-scale 
magnetic fields and the field reversals from arm to arm 
(after Han et al. 2002). 
}\ec 
 
\vs 

The effects of HII regions on 
pulsar DM  have been investigated by Walmsley \& Grewing (1971).
More recently, Mitra et al. (2002) have studied 
in detail 
several   fields along the Galactic plane and showed that not
only 
do the DMs increase but 
the RMs are 
also affected. In
particular, some HII regions were shown to possess magnetic
fields that are 
reversed in 
direction relative to the
surrounding (regular) magnetic field. 


\section{RADIO POLARIZATION OF DIFFUSE EMISSION} 
 
The prediction of Shklovsky (1953) that synchrotron emission
would be linearly polarized led to a number of attempts 
at 
its
detection.
The earliest attempts to observe linearly polarized
radio emission were made by Thomson (1957), Razin (1958) and
Pawsey \& Harting (1960). All these early observers used
small antennas at lower radio frequencies that, in
retrospect, makes polarization detection hardly possible. 
 
\subsection{The Galactic Disk} 
 
Diffuse radio emission comes from the thin disk and thick disk
of our Galaxy (Beuermann et al. 1985), with many spurs 
emerging from the disk (Haslam et al. 1982). The polarized
emission at a few hundred MHz mainly comes from regions
within a few hundred parsecs from the Sun (e.g., Spoelstra
1984) while 
the 
polarized 
emission at higher frequencies from more distant
regions (Junkes et al. 1987; Gaensler et al. 2001). 
 
The first definite detections of polarized Galactic emission
in our Galaxy 
at 408\,MHz 
reported by Westerhout et al. (1962) and Wielebinski et al.
(1962). Both groups had to understand and correct the
instrumental polarization of their radio telescopes which
produced effects larger than the observed radio polarization
signals and showed the ``fan'' structure of polarization
angles around $l=140\dg$, $b=8\dg$. Soon after this
pioneering work several groups took up the challenge to
investigate the distribution of polarized emission in the
Galaxy. The variation in the position angle of the observed
polarization along with the ionospheric data (Wielebinski \&
Shakeshaft 1962) revealed that Faraday rotation was taking
place in the Earth's ionosphere. 
 
The observation of polarized emission at 
higher radio frequencies 
(Muller et al. 1963) showed additional Faraday rotation 
due to 
the magnetic fields in the interstellar medium of the Milky Way. 
Surveys of larger areas of 
linearly polarized radio emission 
were continued at several frequencies (e.g., Berkhuijsen \& Brouw 
1963; Wielebinski \& Shakeshaft 1964; 
Mathewson \& Milne 1965; Berkhuijsen et al.  1964; Mathewson 
et al. 1966; Baker \& Smith 1971; Wilkinson 1973; Baker \& Wilkinson 1974). 
Some observations were dedicated to the large radio features such 
as the North Polar Spur or other loops (Berkhuijsen 1971; 
Spoelstra 1971, 1972a, b, c). Early observations at 1400\,MHz 
by Bingham (1966) showed that polarized emission was widely
distributed in the Galaxy. A milestone in the observation of
Galactic radio polarization 
is 
the map of the northern sky at five
frequencies between 408\,MHz and 1411\,MHz, 
presented by Brouw \& Spoelstra (1976).
After this, mapping of Galactic polarization 
ceased and the 
attention of the observers turned to external galaxies (see the 
next Section) or large-scale continuum high-resolution surveys 
of the Galactic plane at 11\,cm (Reich et al. 1984, 1990; 
F\"urst et al. 1990a, b) and 
to 21 cm
observations. However, recently, 
there is a 
renaissance in
the polarization
measurement 
(Reich  et al. 1990, 1997, 2001; Dickey et al. 
1999; McClure-Griffiths et al. 2001). 
 
An early all-sky map at 150\,MHz was published by Landecker \&
Wielebinski (1970). The all-sky map at 408\,MHz (Haslam et al.
1982) has been available since 1982 after combining the
survey data 
of both northern and southern hemispheres (Haslam
et al. 1974, 1981). Radio maps of large areas of the sky at
other frequencies are also available now (e.g., 22\,MHz by
Roger et al. 1999; 0.2 to 13.8\,MHz by Manning \& Dulk 2001;
45\,MHz by Maeda et al. 1999; 2.3\,GHz by Jonas et al. 1998).
 
The ``Return to the Galaxy'' was heralded by the analysis of 
the polarization data in the $\lambda$11\,cm Galactic plane
survey by Junkes et al. (1987). This higher frequency survey 
showed that, although most of the emission is local, some
polarized radio waves must have come from some more distant
features in the inner Galaxy. This agreed with 
the observations
of the Galactic center  by Seiradakis et al. (1985) at
$\lambda$2.8\,cm 
which showed a complex vertical magnetic field
pattern and by Duncan et al. (1998) on the kpc polarized
plume. A polarization survey of the {\it southern} Galaxy was
made at 2.4\,GHz by Duncan et al. (1995, 1997) using the 
Parkes radio telescope. The reduction of the $|b|<5\dg$ strip
of the 2.7\,GHz {\it northern} survey was made by Duncan et al.
(1999). All the evidence from both these surveys reveals 
a decreased 
polarized intensity 
in the inner plane due to Faraday depolarization.

The mapping of a wide strip of the Galactic plane at 1.4\,GHz, 
called the ``medium latitude survey'', was started in
Effelsberg by Uyaniker et al. (1998, 1999, see Fig.\,5). 
As of date,
large
sections of the Galactic plane have been observed. The aim of
this project is to map the whole 
Galactic plane visible from
Effelsberg within $b =\pm20\dg$. In fact, amazing structures
are seen in polarized intensity in the anticenter regions,
implying very turbulent and quite deep Faraday depolarization
effects.


\vs\vs

\noindent
\psfig{figure=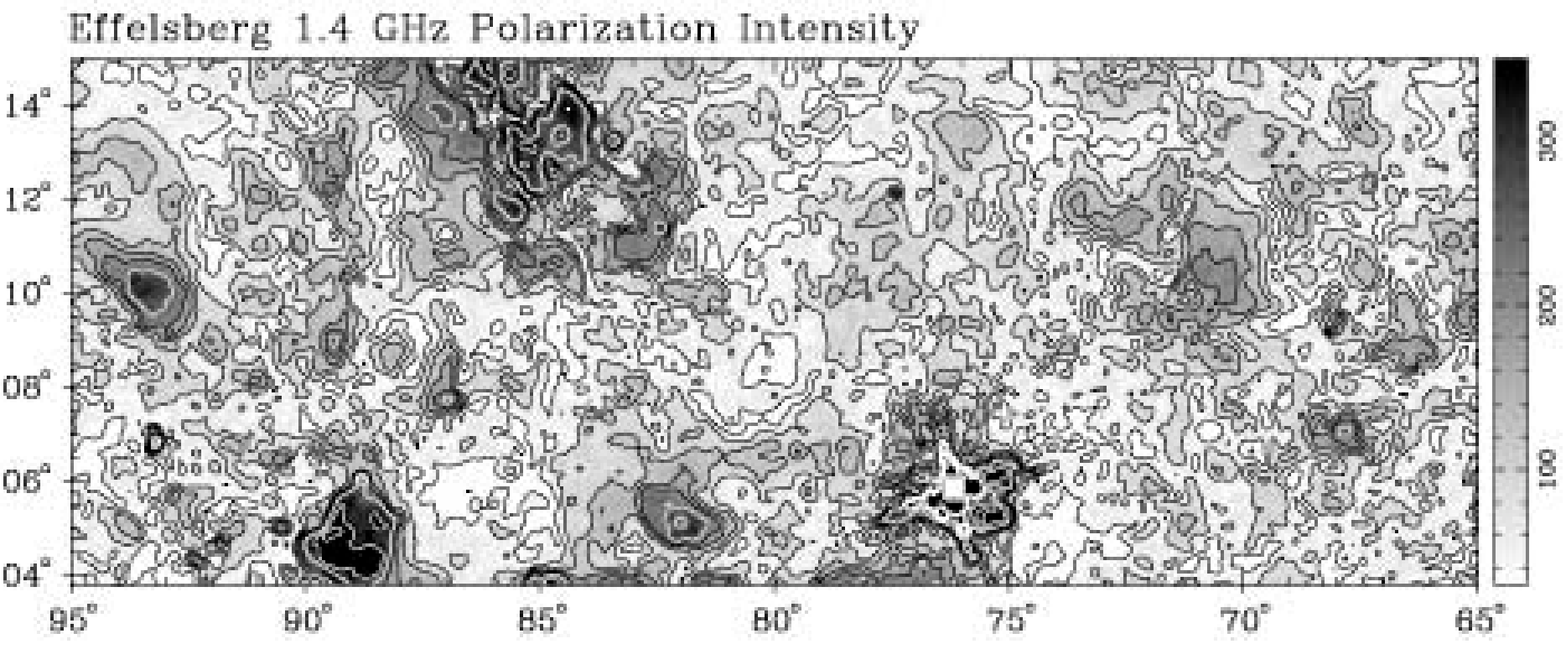,width=55mm,angle=270}\\
\psfig{figure=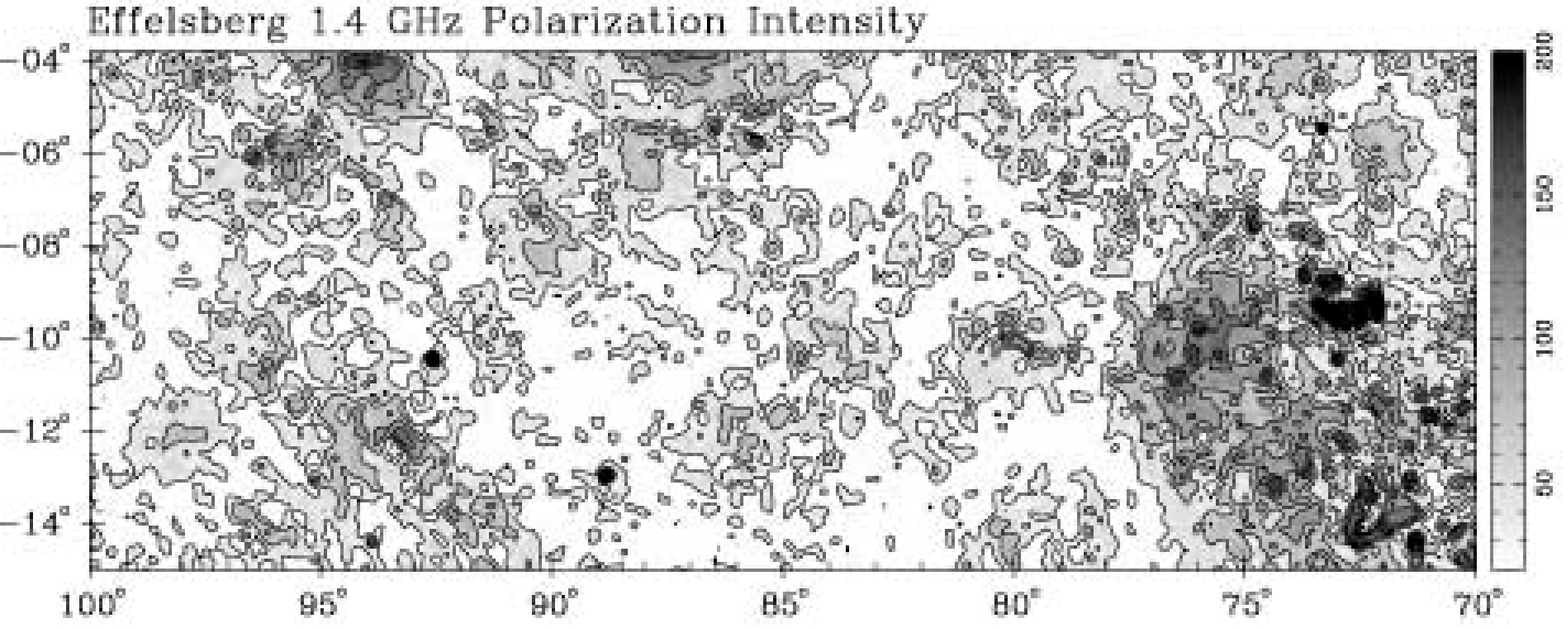,width=55mm,angle=270}

\bc\parbox{125mm}{\small
Fig.\,5~~ 
Polarized intensity map of the outer Galaxy observed by 
   Effelsberg telescope (courtesy W. Reich after Uyan{\i}ker et al.  2001).
}\ec 
 
\vs \vs
 
The observed polarized emission is a modulation of an
intrinsically highly polarized synchrotron background by
Faraday rotation in the diffuse ionized gas in foreground
material, such as H II regions in which the magnetic field is
disordered on scales of ~0.1--0.2\,pc (Gaensler et al. 2001).
In the Canadian Galactic plane survey, Gray et al. (1999)
have identified several new remarkable phenomena around the
W3/W4/W5/HB3 H II region/SNR complex in the Perseus Arm. The
regular features detected in the polarization angle are
superposed on the linearly polarized Galactic synchrotron
background emission by Faraday rotation arising in foreground
ionized gas having an emission measure as low as 1
cm$^{-6}$~pc. The `mottled' polarization arises from random
fluctuations in a magneto-ionic screen of a medium in the
vicinity of the H II regions themselves. At low frequencies,
very complicated patterns are often observed in the polarized
intensity and polarization angles on scales down to a few
arcmin (Wieringa et al. 1993; Haverkorn et al.
2000). Obviously complete depolarization occurs in a very
small depth of the local magnetoionic medium. 
 
The comparison of high-frequency polarization maps of Galactic
regions suggests that 
the RMs may be up to $>$200 rad~m$^{-2}$. 
The polarization observations with arcminute resolution at
20\,cm by Gaensler et al. (2001) in a large section of the 
southern sky show that the RMs 
vary between $\pm$150
rad~m$^{-2}$. Some areas show considerable positive RMs while
others have negative RMs on scales of up to a degree.
Comparing the RM values with those of pulsars, some of the
polarized features at 1.4\,GHz probably originate in the
spiral arm at a distance of a few kpc from the Sun (Dickey
1997). 
These latest results are of great significance because the rest of 
the Galaxy as well as the cosmological sky are observed through this 
foreground magnetic screen. 

\subsection{The Galactic Center} 
 
The Galactic center is the first radio source discovered by 
Carl Guthe Jansky in 1933. In the center there are the strong
poloidal (dipole) fields (e.g., Reich 1994) and toroidal
field (Novak et al. 2002, see Fig.\,6). The best illustration of 
the 
poloidal fields 
is provided by 
the fine, highly polarized radio structures. 
 
The first non-thermal filaments in the Galactic center were
discovered by Yuzef-Zadeh et al. (1984) with VLA
observations. This giant polarized radio arc is very
striking, 
extending more than $1\dg$ on each side of the
Galactic plane (Seiradakis et al.  1985; Tsuboi et al.  1986;
Haynes et al.  1992). 
Later, radio observations using
both single dish radio telescopes (Effelsberg 100\,m, Nobeyama
45\,m) and large synthesis telescopes (VLA, MOST, ATCA), have
revealed many filaments, arcs or loops, threads (e.g.,
Anantharamaiah et al. 1991; LaRosa et al. 2000, 2001; Lang et
al. 1999a, b), plumes (e.g., Duncan et al. 1998) and lobes
(Sofue 1985). Most of these have been studied in great detail
regarding the fine 
structure, polarization at several
frequencies and hence the RMs and spectral 
indices examined.
These strongly polarized 
nonthermal features are very narrow,
with lengths of tens of pc but widths of less than 0.5\,pc
$^*$ \footnote{$^*$ The fractional polarization can be
artificially high (90\%) in 
the VLA observations, due to the
missing total flux at short spacelines by VLA measurements
(see Lang et al. 1999).} (e.g., Inoue et al. 1984; Seiradakis
et al. 1985;  Tsuboi et al. 1986; Haynes et al. 1992).  All
but one are perpendicular to the Galactic plane (Lang et al.
1999b). The intrinsic magnetic field in the filaments or
threads is predominantly aligned parallel to 
the filament (e.g., Lang
et al. 1999a, b; Sofue et al. 1987). They are believed to be
manifestations of strong vertical field lines (mG strength),
i.e., they are illuminated flux tubes in large-scale 
pervasive fields (Yusef-Zadeh \& Morris 1987a; Uchida et al.
1996).  A consensus reached now is that these must be the
consequence of a substantial poloidal magnetic field which
pervades the central $\sim$100\,pc of our Galaxy (Yusef-Zadeh
\& Morris 1987a; Morris \& Yusef-Zadeh 1989). A primordial
origin is suggested for the poloidal field.  Equipartition
magnetic field values in the magnetic tubes are at least 70
$\mu$G, perhaps about 0.4\,mG (Bicknell \& Li 2001) or even
several mG (Yusef-Zadeh \& Morris 1987b). 
 
Large RMs have been observed with irregular variations along
the filaments (Inoue et al.  1984; Seiradakis et al. 1985; 
Sofue et al. 1987). The Faraday screen is very probably close
to the Galactic center. 
Within a few arcminutes, the  RM value
can 
vary from 500 to $-$500 rad~m$^{-2}$ in ``the Pelican''
(Lang et al. 1999a), 
from 100 to 2300 rad~m$^{-2}$ in the Northern
Threads (Lang et al. 1999b), 
from 2000 to 5500 rad~m$^{-2}$ along
``the Snake'' (Gray et al. 1995), or from $-$4200 to $-$370
rad~m$^{-2}$ in 
the filament G359.54+0.18 (Yusef-Zadeh et al. 
1997). In the 
southern plume of the largest filament, the RM
values are all negative with a maximum up to $-$2500
rad~m$^{-2}$, but in the 
northern plume 
the RMs are mostly positive
with some negative holes, the positive values up to 1000
rad~m$^{-2}$ (Tsuboi et al. 1986). No coherent structure 
in the 
RMs can be found. However, Novak et al. (2001) noticed that
the 
dominant RM values in the different ($l,b$) quadrants have an
antisymmetric distribution, which probably indicates that in
the whole Galactic halo from the Galactic center, the toroidal
fields have different directions above and below the Galactic
plane (see Han 2002).

\vs\vs

\psfig{figure=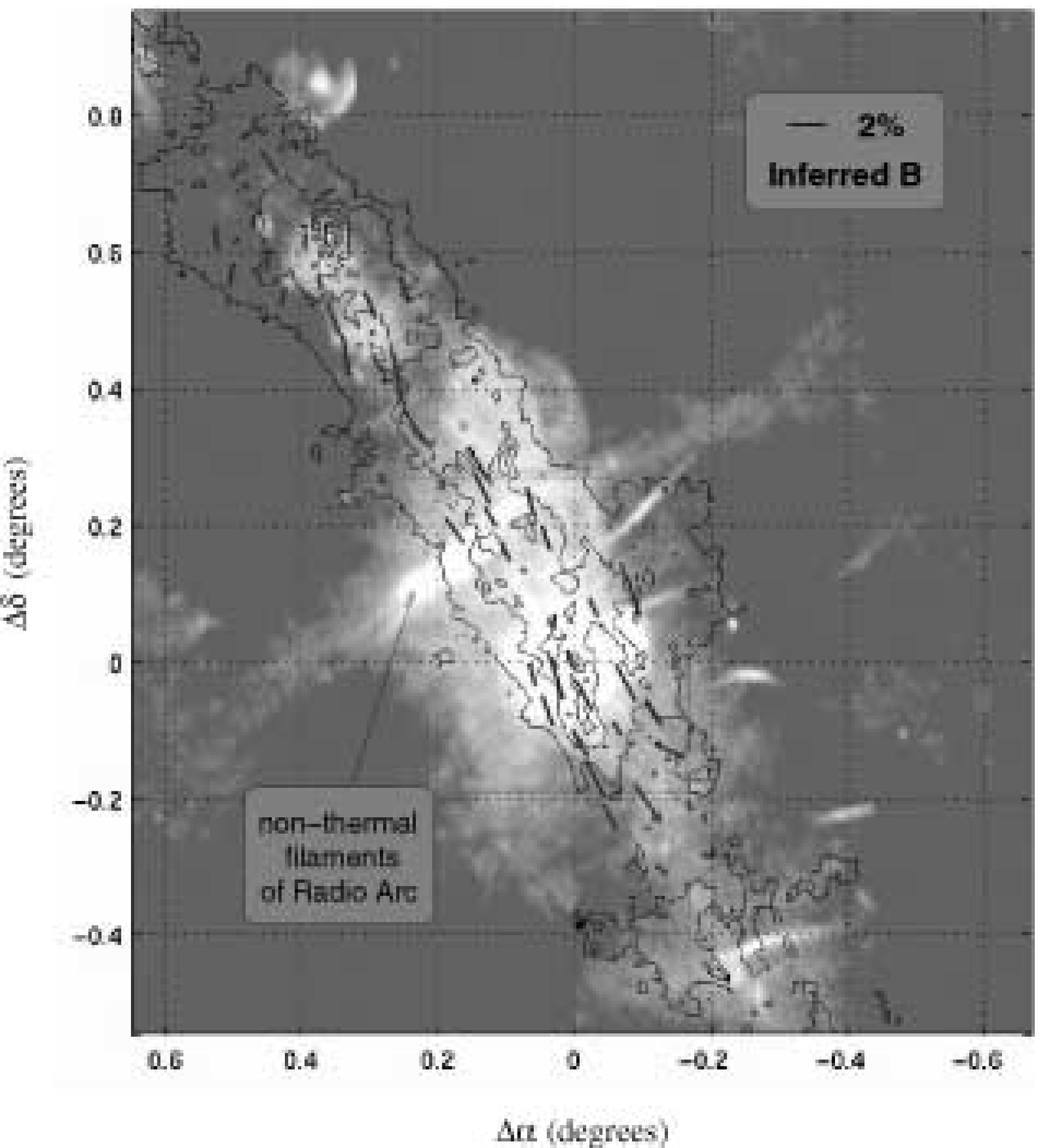, height=115mm}
\vs
\bc\parbox{120mm}{\small
Fig.\,6~~ 
Radio structure of the Galactic center at 90\,cm and the 
toroidal magnetic fields in the central molecular zone revealed by the 
polarization at a submillimeter band (courtesy N.E. Kassim \& D.T. Chuss. 
See Novak et al. (2002) for other information).
}\ec 
 \vs

Though the connection of the filaments and the molecular
clouds or HII regions 
can be identified 
in the observations
(e.g., Uchida et al. 1996; Yusef-Zadeh et al. 1997; Reich et
al. 2000), and magnetic fields can be anchored in 
the molecular
clouds (Bicknell \& Li 2001), the detailed process for
particle 
acceleration is still a puzzle. It is not clear
where the relativistic particles come from for the
synchrotron emissions and why these 
particular field lines are 
picked out 
for illumination. 
 
The observations of the Galactic center at mm, sub-mm and
far-infrared bands have revealed the toroidal magnetic fields
in the central molecular cloud-zones (Novak et al. 2000,
2002) parallel to the Galactic plane, as discussed earlier. 
 
\subsection{Nearby Galaxies} 
 
The first report of the presence of polarized radio emission
(and hence magnetic fields)  in an external galaxy came from
Mathewson et al. (1972) using the 
then newly completed Westerbork
Synthesis Radio Telescope. In
this early work, the orientation of the {\bm E} vectors could
be shown to be in agreement with the optical observations of
Appenzeller (1967) for M51. This  work was followed up
by Segalovitz et al. (1976) who mapped M51 at a second
frequency as well as tried  to observe M31.

Observations by Beck et al. (1978, 1980)  of M31 using the
100\,m Effelsberg dish at $\lambda=$ 11\,cm allowed 
a definite
detection of 
magnetic field,
which 
showed a surprisingly
homogenous distribution. From that time onwards practically
all large nearby galaxies 
have
been mapped, at ever higher
frequencies, first in Effelsberg and later with the VLA (see
reviews by: Wielebinski \& Krause 1993; Beck et al. 1996;
Beck 2000). An MPIfR polarimeter was taken to the Parkes
radio telescope and led to the observations at several
frequencies of the Magellanic Clouds (Haynes et al. 1986,
1991). The magnetic field in the Large Magellanic Cloud
resembles a trailing spiral pattern around the kinematical
center (Klein et al. 1993). 
 
Multi-frequency 
maps, when made with identical angular
resolution, allow 
a determination of the Faraday rotation
in the spiral arms of 
the galaxy as well as the  orientation of
the original {\bm E} vector and hence of the inherent
magnetic field. The magnetic fields were found to be
orientated with the spiral arms on a very large scale
(kiloparsecs) and confined to the galactic plane in most
nearby grand-designed spiral galaxies, such as M51 (Neininger
\& Horellou 1996), M81 (Krause et al. 1989b), M83 (Neininger
et al. 1993) and NGC~2997 (Han et al. 1997). 
 
A long discussion was begun, pushed by the theoretical
considerations, 
on the 
origin of the  cosmic magnetic
fields (e.g., Tosa \& Fujimoto 1978; Beck et al. 1996), with
the aim of determining if the magnetic fields in galaxies
are bi-symmetrical (BSS) or axi-symmetrical (ASS). The
question of field configurations was repeatedly asked by the
two diverging factions of theoreticians, the dynamo community
wanting to see rings or ASS fields, while the primordial
origin people wanting to see BSS fields. The variation of 
the RM 
as a function of the azimuthal angles has been used to
distinguish between the BSS field and the ASS or ring field
structure (see e.g., Krause 1989a, b). Detailed analysis of
the 
RM distribution of several galaxies, though with large
measuring 
uncertainties, showed that many galaxies have a
strong ASS field as well as a BSS field. In some galaxies, the
ASS field 
dominates (Krause \& Beck 1999), while in others a
BSS field seems to be dominant (M81, M51,  see detailed
situations in Table 1 of Beck 2000). Multi-mode fits to 
BSS 
and ASS patterns became necessary to interpret the
detailed observational data. One should note, however, that
the observed polarized emission mainly 
comes from a
thinner disk, while the RMs are caused by the electrons and
magnetic fields in the closer layer of a thicker disk or halo
(plus the thinner disk). In some cases 
the halo contribution may
be significant (e.g., Soida et al. 2001). Observations of
irregular 
dwarf galaxies without systematic gas motions
(Chyzy et al. 2000) and flocculent galaxies with only
rudimentary spiral arms (Knapik et al. 2000) also revealed
well-organized 
magnetic fields, indicating 
that some
non-standard dynamo or other mechanisms 
are responsible for the regular magnetic fields.

Higher resolution observations of some galaxies (e.g.,
IC342: Krause 1993; M83: Beck 2000; NGC~2997: Han et al.
1999, see Fig.\,7;  NGC~1097: Beck et al. 1999; M51: Neininger \& Horellou
1996) revealed that the polarized emission is strongest in
the inner edge of the optical spiral arms or along the dust
lanes, clearly associated with local compression effects by
shock. This mostly happens in the inner disk ($<2/3$ optical
radius). However, an important result was the discovery of
``magnetic arms'' in the outer disk. These aligned magnetic
fields exist between the optical spiral arms or extend
independently from optical arms in some galaxies, e.g., 
NGC~6946 (Beck \& Hoernes 1996), NGC~2997 (Han et al. 1999) and
IC342 (Krause et al. 1989).

Recent developments coming from the polarization observations
of barred galaxies, NGC~1097 (Beck et al. 1999), showed  the
magnetic fields following the gas streamlines and aligned with
shearing flow. The observations also probed the magnetic field
near the circumnuclear disk -- a new area to explore in the
future.

\vs\vs
\psfig{figure=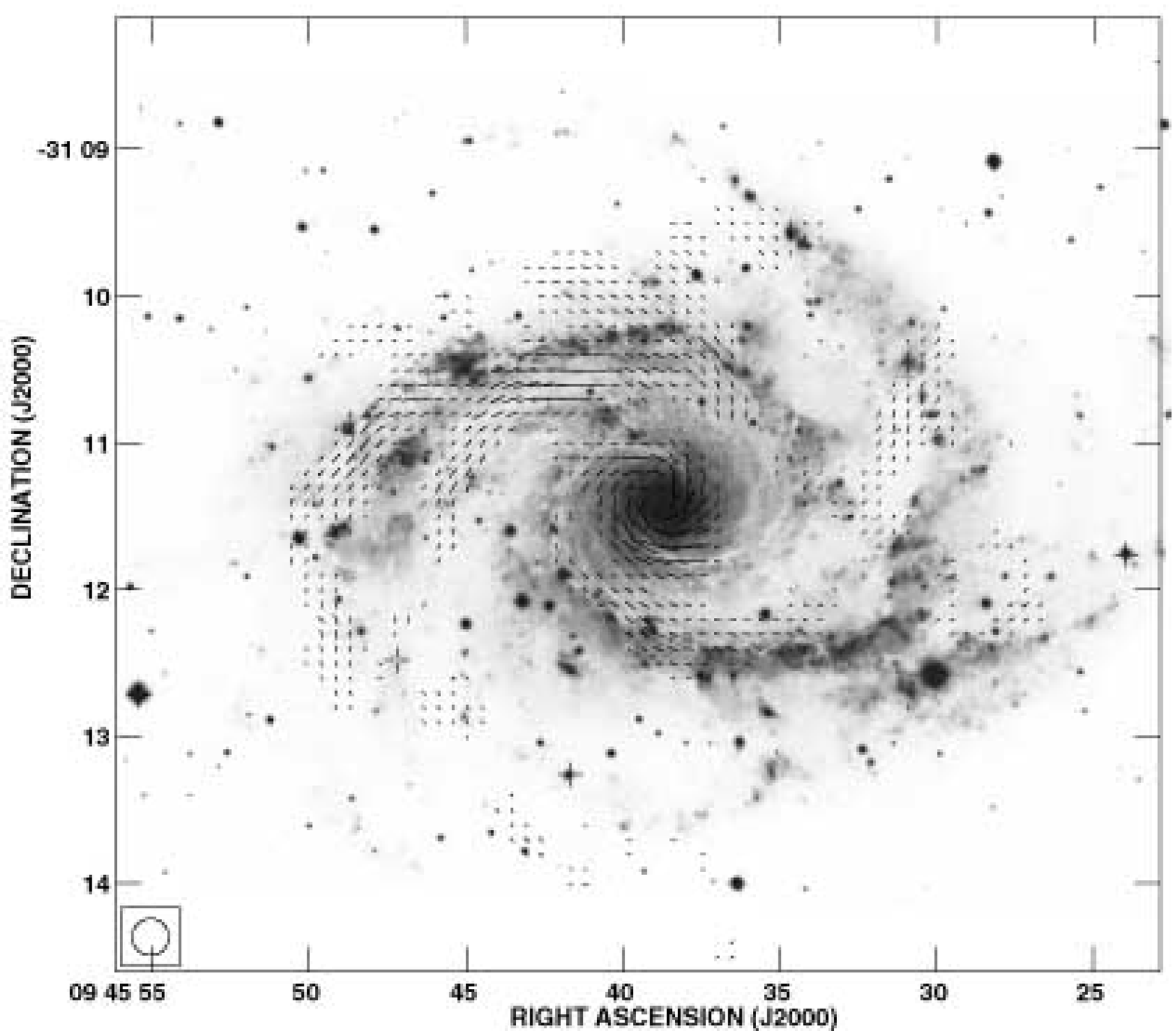, height=115mm}
\hspace{115mm}
\bc\parbox{120mm}{\small
Fig.\,7~~ 
Polarization observation of 
the grand-design spiral galaxy 
NGC~2997 at 6\,cm by VLA (after Han et al. 1999). 
The 
polarization vectors $E$ were rotated 
by 90$\dg$ to show the 
$B$ orientations.
}\ec

\vs \vs

Extended radio halos have been observed 
in several galaxies
but only a few of them have ordered polarization 
reflecting 
ordered magnetic fields (e.g., Dumke et al.
1995). In some galaxies, like NGC~4631, M82  and the Milky 
Way, vertical or poloidal magnetic fields shown by polarized
spurs 
are observed in the nuclear area (e.g Golla \& Hummel 
1994; Reuter et al. 1994). The magnetic fields in the halo 
probably result from outflows of gas from the disk into the
halo and the cosmic rays for the radiation probably come from
some 
central starburst.  NGC~4666 may be another example of
such 
starburst-driven superwinds (Dahlem et al.
1997). 
 
\subsection{Radio Galaxies and Quasars} 

Radio galaxies and quasars in the distant universe are most
intensively illuminated by synchrotron emission of the
relativistic particles in magnetic fields, both in the
energetic jets of the central blackhole or in the outer lobes
which are produced by the interaction of the jets with the
intergalactic medium. The central core component is
physically associated with galactic nuclei but can be
resolved into 
finer structures 
when observed with VLBI. 
 
Polarization studies of radio galaxies and quasars started 
with the detection of extended polarized components and their
rotation measures in the nearest radio galaxy -- Centaurus A 
(Cooper \& Price 1962). Already at that time it was shown
that the radio polarization observations of Centaurus A 
agreed well with 
the optical polarization studies by Elvius and
Hall (1964). Radio polarization studies afterwards have 
demonstrated that the projected magnetic fields can appear
uniform on scales exceeding a hundred kpc, or sometimes over
several hundred kpc! 
 
In the intervening years, hundreds of polarization 
maps of radio
galaxies have been published, first at lower frequencies and
now at higher frequencies and with better resolutions mostly
by VLA and WSRT observations (e.g., H\"ogbom \& Carlsson
1974; Klein et al. 1994; Ishwara-Chandra et al. 1998). Many
objects have multifrequency polarization data 
that have 
been 
analysed 
for magnetic fields (e.g., Johnson et al. 1995; Murgia et
al. 2001). The fields are reflected by the polarized radio
emission and their intrinsic direction (orientation). If
corrected by foreground RMs, the fields are generally
circumferential (see Fig.\,8), 
either
parallel with (for FR-II and one-side jets in
FR-I sources) or perpendicular to (for two-sided jets of FR-I
sources) the elongated jet or bridge component; in the lobes it
wraps around the source edges or strong intensity gradients
(see reviews by Miley 1980; Bridle \& Perley 1984; Saikia \&
Salter 1988).

The magnetic field is irregular but sheared and compressed to
a preferred direction so that the polarized radio emission is
observed following the jets or the bands of jets. The 
deceleration or transverse shocks lead to compression and an
orthogonal field. Due to the high intensity, radio mapping at
arcsec resolution scale has been possible. A result of great
significance is the ``jump'' of polarization vectors (and
hence of magnetic field) by 90$^{\circ}$ from the center to
the end of jets in some sources (see Bridle \& Perley 1984).
This seems to be a method of 
stabilizing 
the radio
emission. It could be due to the action of a helical magnetic
field in these cases. 
 
Many objects (or even core components) have been studied using
high resolution multifrequency  VLBI/VLBA polarization data
(e.g., Venturi \& Taylor 1999). The most recent development
is the parsec-scale RM project using the VLBA (Taylor 2000)
to test the unified models for AGN by looking for orientation
effects in 
the RM distribution of 
the quasar core. Probably the
clouds entrained by jets and magnetic fields are responsible
for the observed large variations of the RMs at pc or sub-pc
scales (Zavala \& Taylor 2002). 
 
Detection of ordered magnetic fields at scale of hundreds to
thousand pc in high redshift objects of $z>2$ (Udomprasert et
al. 1997; Athreya et al. 1998) poses a stiff challenge to the
dynamo mechanism for generating magnetic fields, primarily
due to the very short time available for amplification of the
initial seed field. 
 
Recently, the detection of {\it circular polarization} 
in 
AGN cores and nearby galaxy 
nuclei by the VLBA and ATCA (Homan et
al. 2001; Rayner et al. 2000; Bower et al. 1999; Brunthaler
et al. 2001) re-opened 
a window 
on the physical
conditions and magnetic fields in the core regions. 
Typically, the mean degree of circular polarization is less
than 1\%.  Its origin is not yet understood. However, a few
possibilities exist such as synchrotron emission, cyclotron
emission, coherent emission mechanisms, Faraday conversion in
relativistic plasma and Scintillation-induced circular
polarization (e.g., Macquart 2002). The 
radiation transfer processes 
seem to be the most promising.

\vs\vs\vs\vs\vs

\hfill\psfig{figure=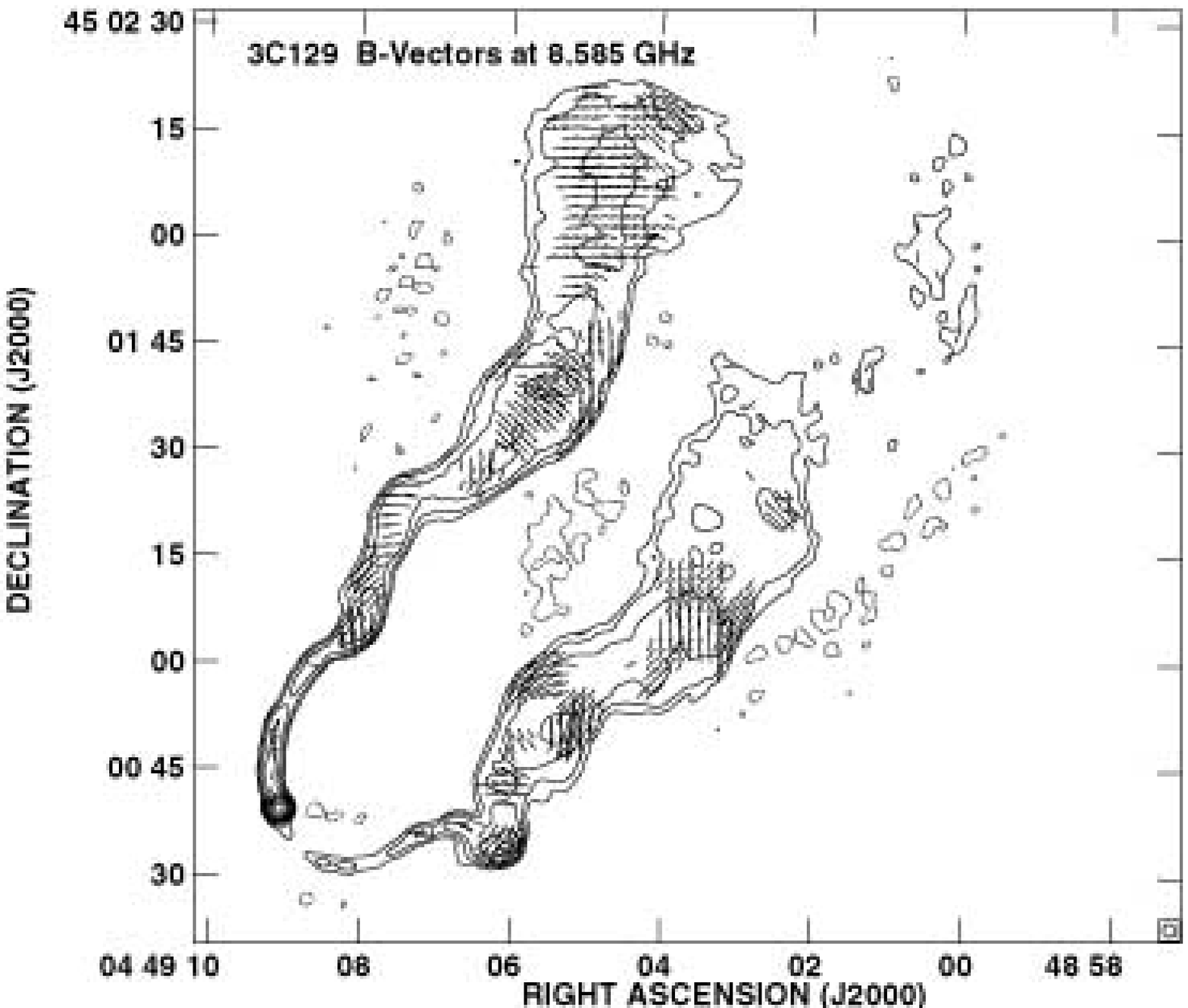, height=85mm} \hfill~~~

\bc\parbox{110mm}{\small
Fig.\,8~~ 
Magnetic field vectors in the radio galaxies 3C 129 
in the 3C 129 cluster (8.4\,GHz observations,  RM-corrected). 
(after Taylor et al. 2001).
}\ec 
 
 \vs\vs

\subsection{Clusters of Galaxies} 
 
Diffuse radio emission in clusters of galaxies is due to
intracluster magnetic fields. The estimation of the magnetic
fields were mostly made through observations of radio halos
and variable RMs of the sources in the clusters or 
the RMs of the
background sources, though the lower limit of the field
strength can be estimated from the hard X-ray detection. The
most recent review on this topic by Carilli \& Taylor (2002)
has included a very good summary of recent progress in all
respects.
 
A report of a halo in the Perseus cluster of galaxies was
made by Ryle \& Windram (1968), but the first clear detection
was actually made by Large et al. (1959) in the Coma cluster.
This was confirmed by 
the many studies 
that followed (e.g.,  Willson
1970; Wielebinski 1978).  Further halos were found in the
A754 (Wielebinski et al. 1977), A1367 (Gavazzi 1978) and
other clusters A2255, A2256 and A2319 (e.g., Harris et al.
1980; Bridle et al. 1979; Harris \& Miley 1978) though many
candidates were reported (see the status 
presented by Hanisch
1982).  Using recent radio survey data which cover most of
the
known clusters, a large sample of halo candidates has been
found (e.g., Kempner \& Sarazin 2001; Giovannini et al. 1999)
and some of them have been confirmed already (e.g., Govoni et
al. 2001a). Up to now, more than a dozen of halos are
certainly detected and the number is increasing, maybe up to
more than two dozen soon. Evidently, almost all halo clusters
present clear features of recent mergers (e.g., Markevitch \&
Vikhlinin 2001; Feretti et al. 2001), indicating that the
nonrelaxed state in the cluster is related to the radio
emission. 
A strong correlation between 
the X-ray luminosity and
halo radio power 
has been found (Liang et al. 2000; Govoni et al. 2001a). 
 
The spectrum of the Coma halo emission has been a subject of
numerous studies (e.g., Schlickeiser et al. 1987; Giovannini
et al. 1993) 
which confirmed the non-thermal nature of the halo
and hence 
its magnetic origin. The spectrum of 
the halos is very
steep, often about $-$1.5 (Deiss  et al. 1997; Liang et al. 
2000), see 
the list in Henriksen (1998).
 
Not much radio polarization has been detected from the halo,
except for some polarization associated with the halo-`G' in
A2256 (Bridle et al. 1979), while multifrequency polarization
observations revealed the polarization of many cluster
objects (e.g., Feretti et al.  1995; Taylor et al. 2001;
Govoni et al. 2001; Eilek \& Owen 2002). Analyses of both the
distribution and the variance of RMs of extended radio
galaxies in the cluster (e.g., Dreher et al. 1987; Taylor et
al.  2001; Govoni et al. 2001b; Eilek \& Owen 2002) or the
background sources (Vallee et al. 1987; Kim et al. 1991;
Clarke et al. 2001) have almost reached 
the following 
consensus: the
random magnetic field in the intracluster gas has a strength
of 5\,$\mu$G at a length scale of about 10\,kpc, within a
factor of 2 or 3 for different authors or different 
objects/regions. The magnetic field could be stronger towards
the cluster center (e.g., Govoni et al. 2001b) and very high 
RMs (of radio galaxies) have been found near the center of the
cooling flow clusters (e.g., Ge \& Owen 1993). Recently, it
has been found that X-ray emission is well correlated with
the rms 
value of the cluster rotation measures (Dolag et al. 2001),
implying that the the field strength 
decreases with 
the radius
according to $B\sim N_e^{0.9}$ for A119. The central 
density-enhanced ``cooling flow'' regions can have magnetic
fields up  to $\sim$40\,$\mu$G with coherence scales up to
$\sim$50\,kpc (Taylor \&  Perley 1993). 
 
Radio relic sources have been detected in or between clusters of
galaxies, for example, the bridge emission between Coma and A1367
(Kim et al. 1989), and the extended source adjacent to three Abell
clusters of galaxies of Rood Group \#27 (Harris et al. 1993).  All
relic sources have low surface brightness and obviously reflect the
magnetic fields between the clusters. Some relic sources are significantly
polarized.

\section{THE ORIGIN OF MAGNETIC FIELDS} 
 
We have no direct observational evidence for the first
cosmological magnetic fields. The fields, with observed
intensities of $B\sim 2 - 10\,\mu$\,Gauss today must have
started at a lower level and become amplified by some
mechanism at 
some cosmological stage (cf. Kulsrud 1999).  The
``battery effect", proposed by Biermann (1952), generates
magnetic fields of $B\sim 10^{-20}$\,G at the most (e.g.,
Lesch \& Chiba 1995). Several mechanisms to produce the seed
fields at different epochs of the universe have been
proposed, for example, Harrison (1973) aregued that
turbublence in the radiation-dominated era produces the weak
seed fields of presently $10^{-8}$~G; Hogan (1983) gave a
basic argument 
for 
considering phase 
transition as 
a potential
mechanism 
for the generation of primordial magnetic fields;
Turner and Widrow (1988) proposed an inflation scenario for
the creation of primordial magnetic fields; Quashnock et al.
(1989) considered the thermoelectric effect in QCD phase
transition to generate the magnetic field of
$2\times10^{-17}$~G in 
the early universe; Wiechan et al. (1998)
showed that the 
plasma-neutral gas friction in a weakly
ionized rotating protogalactic system creates the seed
magnetic fields.  The Biermann battery in ionization fronts
can in principle produce reasonably strong magnetic fields 
(see Subramanian et al. 1994; Gnedin et al. 2000).  
Magnetic fields could originate deep in the early phases of
the universe (see Grasso \& Rubinstein 2001 for a review). 
The role of magnetic fields in the very early Universe, the
symmetry braking during phase transitions or at the time of
recombination (e.g.,  Sicitte 1997),  structure formation
(e.g., Totani 1999; De Araujo \& Opher 1997; Battaner et al. 
1997)  even  the big-bang nucleosynthesis, should be fully
investigated (See Rees 1987).
 
Some progress has been made recently in the studies of the
evolution of such primordial magnetic fields (e.g., Kulsrud
et al. 1997; Howard \& Kulsrud 1997). Some early papers
(e.g., Piddington 1964, 1978) considered that if an
intergalactic magnetic field exists with $B\sim 10^{-8}$~G,
it could be compressed by galactic rotation to
$B\sim10^{-5}$~G, 
intensities that are actually observed in 
galaxies. This was supported by some 
papers (e.g.,
Sofue et al.  1979; Anchordoqui \& Goldberg 2002) that
claimed the detection 
of 
a lower limit of 
intergalactic
magnetic field with $B$ above 3$\times 10^{-9}$~G. However,
these results contradict others (e.g., Blasi et al. 1999). 
Such a primordial magnetic field could in principle be
detected by the Faraday rotation of microwave background
polarization (Kosowsky \& Loeb 1996) or by the correlation of
the Faraday sky (Kolatt 1998).
 
If only 
a small seed magnetic field was generated in the
early universe, then a rather large amplification, by a
dynamo, for example, is needed to reach the observed values
in cosmic objects (e.g., Parker 1979). However, magnetic
fields in high redshift ($z\sim$ 2.0 -- 3.0) objects with
Lyman-$\alpha$ damped systems have been found 
to be 
as strong as in
nearby galaxies and 
in some clusters of galaxies (e.g., Kronberg
et al. 1990, 1992; Athreya et al. 1998). This poses some
challenge to galactic dynamo, since, at that time, galaxies
should have rotated only a few times.  The studies on the
variance of residual 
RM (observed 
RM minus foreground
Galactic 
RM) versus source redshift $z$ (e.g., Oren \& Wolfe
1995; Perry et al. 1993), mainly due to data quality, have
produced controversial results on the cosmological evolution
of the magnetic field (in the intervening clouds). At
present, evidence has been accumulated for magnetic fields in
the intergalactic space.  In fact, magnetic fields can be
easily-developed in the vicinity of an accreting disk of the
black hole in the galaxy or quasar center (e.g.,  Contopoulos
\& Kazanas 1998), and then their outflows can magnetize a
large fraction of the entire intergalactic medium (e.g., Daly
\& Loeb 1990; Furlanetto \& Loeb 2001; Kronberg et al.
2001).  Furthermore, the superwinds of primeval galaxies can
also effectively seed  magnetic fields into the intergalactic
medium (Kronberg et al. 1999; Birk et al. 2000).

The explanation of the origin of Earth's magnetic field was the
triumph of the mean field dynamo theory as developed by Parker
(1955) and 
Steenbeck \& Krause (1969a).
This theory could be applied to planets (Steenbeck \& Krause
1969b), to magnetic stars , and more recently to spiral
galaxies (Ruzmaikin et al. 1988).  Good reviews of the dynamo
theory can be found in Parker (1979) and Krause \& R\"adler
(1980).  In the dynamo theory, Maxwell's equations are solved
by introducing an $\alpha$ term that describes ``looping'' of
the magnetic field in the $z$ direction relative to the
$\omega$ rotation.  The solution for the ``$\alpha-\omega$
dynamo'' has been very successful in explaining many
observational results. Stix (1975) applied the dynamo theory
to an oblate spheroid, representing an edge-on galaxy. A
series of papers by Ruzmaikin, Shukurov and Sokolov (see 
the 1988 book) 
have developed the mean dynamo theory to be
applicable to magnetic fields in
galaxies.  In particular, the multi-mode solution (e.g.,
Baryshnikova et al. 1987) has shown that spiral magnetic
fields seem to be possible within the frame of the dynamo
theory.  Most recently detailed discussion on kinematic
dynamo and small-scale field were made by Schekochihin et al.
(2002). The time evolution of a dynamo has been actively
studied by, e.g., Brandenburg et al.  (1989). In fact, a
turbulent dynamo probably could produce strong magnetic
rope-like structure but is difficult to generate large-scale
magnetic fields (Subramanian 1998).  
A large-scale kinetic
helicity is probably 
a necessary ingredient for 
a large-scale
magnetic field (Blackman \& Chou 1997) and helical turbulence
may provide the key to the generation of large-scale magnetic
fields (Blackman 2000; Brandenburg 2001). The density wave in
spiral galaxies can provide a spiral shock wave and the
subsequent $\alpha$-effect which may alter the field
configuration of 
the dynamo (Mestel \& Subramanian 1991;
Subramanian \& Mestel 1993).  Kulsrud (1999) looked into the
many limitations of the dynamo theory.  However, the current
opinion is that most of the phenomena that are observed can
be explained by a modified dynamo theory.  A review of very
recent developments in 
the magnetic dynamo theory, most
impressively on the magnetic helicity and the evolution of
kinetic and magnetic energy spectra, can be found in Blackman
(2002).  Many groups have added different physical
considerations into 
the dynamo and used computer simulations to
show 
possible consequences comparable to observational
results, e.g., 
the bar-effect by Moss et al. (2001) and
Otmianowska et al. (2002), 
the swing-excitation by Rohde et al.
(1999), 
the very detailed considerations of supernova and
superbubble-driven dynamo by Ferri\'ere \& Schmitt (2000) 
and the 
$\alpha$-effect by magnetic buoyancy by Moss et al. (1999).  
 
To explain the magnetic fields in young, high redshift
galaxies, the evolution of protogalactic clouds and
electrodynamic processes (Lesch \& Chiba 1995) or the
cross-helicity effect (Brandenburg \& Urpin 1998) should be
considered. It seems quite possible that large-scale magnetic
fields in galaxies are the residue of dynamo processes in the
plasma before the protogalactic plasma cloud collapsed and
the galaxy 
formed. At that time all 
the conditions are very
suitable for the dynamo operations and 
a seed field can be
easily produced by the Biermann battery -- the large-scale
fields is then of primordial origin (Kulsrud et al. 1997, see
also discussion in Schekochihin et al.  2002). Such a field,
however, has to be maintained by 
the dynamo process in galaxies
after the galaxy was formed.

\section{CLOSING REMARKS} 
 
The subject of magnetic fields is in its infancy. While
gravity (density wave theory) or star formation have received
considerable attention, with large groups working in these
areas, only a few workers have struggled to reveal the
importance of magnetic fields in the 
cosmos. 
Very often the question is asked if magnetic fields are a
basic parameter of our universe or only a consequence of
rotation.  An argument often used is that the energy density
in magnetic fields is much less than in gravitational effects
and is hence unimportant. However, magnetic fields have a
unique directional effect on matter, if coupled. The coupling
of magnetic fields to matter is poorly understood. The
traditional description of ``frozen-in magnetic fields'' must
give 
way to studies of the magnetic properties of
interstellar matter at different temperatures.  For instance
recent sub-mm observations (Wielebinski et al. 1999) have
shown that CO gas in galaxies can be seen with temperatures
of 50\,K or more. Now that we know that ionization in warm
parts of molecular clouds is possible, we can imagine that
magnetic fields could exert sufficient force to be of dynamic
consequence.  A quotation of Lou Woltjer that is often cited
runs: ``the poorer our knowledge the stronger the magnetic
field''.  However, in view of the fact that magnetic fields
are found everywhere in the universe, they deserve
appropriate treatment and study.

\vs\vs\no {\bf Acknowledgements}~~ We thank Dr. Peter W.
Draper (durham.ac.uk) for his re-make of the polarization
picture of NGC~1068 for us, Dr. Rainer Beck and Ms. Catherine
Boersma \& Dr. Melanie Johnston-Hollitt for careful reading
of the manuscript and many corrections. We thank the support
in the partner group frame between Max-Planck Society and
Chinese Academy of Sciences for a long-term cooperation
between the MPIfR and NAOC.  The research of JLH in China is
supported by the National Natural Science Foundation of China
(10025313 and 19903003) and the National Key Basic Research
Science Foundation of China (NKBRSF G19990752).  This review
has made extensive use of NASA's Astrophysics Data System
Abstract Service.

\bibliographystyle{apj1} 
\bibliography{mag_field,psrrefs} 
 
\end{document}